\documentclass{aa}  
\def\lsi{LSI+61$^{\circ}$303}
  
\def\kms{km~s$^{-1}$}  

\def\vsi{$v\: \sin i$}  
\def\rsun{R$_{\odot}$} 
 
\def\msun{M$_{\odot}$} 
\def\Ha{H$\alpha$}

\usepackage{graphicx}
\usepackage{txfonts}
\begin{document}

\title{\Ha\ observations of the $\gamma$-ray-emitting Be/X-ray binary LSI+61$^0$303:
   orbital modulation, disk truncation, and long-term variability  \thanks{based on observations obtained  at the National Astronomical 
Observatory Rozhen, Bulgaria }  }
\titlerunning{LSI+61$^0$303: \Ha\ emission line}
\authorrunning{Zamanov et al. }
   \author{R. Zamanov\inst{1}
          \and
          K. Stoyanov\inst{1}  
	  \and   J. Mart{\'{\i}}\inst{2}	
	  \and   N. A. Tomov\inst{1} 
	  \and   G. Belcheva\inst{3,4}	
	  \and   P. L. Luque-Escamilla\inst{5}  
	  \and   G. Latev\inst{1}
          }  
   \institute{Institute of Astronomy and National Astronomical Observatory, 
              Bulgarian Academy of Sciences, 72 Tsarighradsko Shousse Blvd., 1784 Sofia, Bulgaria
         \\   \email{rkz@astro.bas.bg,  kstoyanov@astro.bas.bg,  jmarti@ujaen.es, tomov@astro.bas.bg}
	 \and
           Departamento de F\'isica (EPSJ), Universidad de Ja\'en, Campus Las Lagunillas,  A3-420, 23071, Ja\'en, Spain 
 	 \and
	  Department of Astronomy, St. Kliment Ohridski University of Sofia, 5 James Bourchier Blvd., 1164 Sofia, Bulgaria 
	 \and
	   Department of Physics and Astronomy, Pevensey II Building, University of Sussex, Falmer, Brighton BN1 9QH, UK
         \and
           Departamento de Ingenier\'ia  Mec\'anica y Minera (EPSJ), Universidad de Ja\'en, Campus Las Lagunillas A3-008, 23071, Ja\'en, Spain 
	   \\
             }

   \date{Received May 30, 2013; accepted  September 12, 2013}
 \abstract{We report  138 spectral observations of the \Ha\ emission line of 
    the radio- and  $\gamma$-ray-emitting  Be/X-ray binary   \lsi\
    obtained during  the period  of September 1998 -- January 2013. 
    From measuring various \Ha\  parameters, we found that  
    the orbital modulation of the \Ha\ is best visible  in  the equivalent width ratio  $EW(B)/EW(R)$, 
    the equivalent width of the blue hump,  and in the radial velocity of the central dip.  
    The periodogram analysis confirmed that the \Ha\ emission is modulated with the orbital and superorbital periods. 
    For the past 20 years the radius of the circumstellar disk is similar to the Roche lobe size at the periastron. 
    It is probably truncated by a 6:1 resonance.  
    The orbital maximum of the equivalent width of \Ha\ emission peaks after the periastron and 
    coincides on average with the  X-ray and $\gamma$-ray maxima.  \\
    All the spectra  are available upon request from the authors and through the CDS.
  }
   \maketitle
%

\section{Introduction}  
The Be/X-ray binary \lsi\  was  first detected  as strong  $\gamma$-ray source  with COS~B (Hermsen et al. 1977) and
as an X-ray source with the $Einstein$  observatory (Bignami et al. 1981).
The Be/X-ray binaries (BeXRB) are systems that consist of a compact object that orbits
an optical companion.   The optical companion is a Be star. Be stars are non-supergiant fast-rotating B-type and 
luminosity class III-V stars that at some point of their lives have shown spectral lines in emission
(Porter \& Rivinius 2003; Balona 2000; Slettebak 1988). The best-studied lines are those of hydrogen (Balmer and Paschen series),
but  the Be stars can also show He and Fe in emission (see Hanuschik 1996 and references therein). 
They also show some  infrared excess. The origin of the emission lines and infrared excess in BeXBR 
is attributed to an equatorial disk that is fed from material expelled from the rapidly rotating Be star. During periastron, 
the compact object passes close to this disk, sometimes may even go through it and cause a major disruption. 
A strong flow of matter is then captured by  the compact object. In  most cases the compact object is a neutron star (NS) 
detected as an X-ray pulsar (Bildsten et al. 1997 and references therein). The conversion of the kinetic energy of the infalling matter into radiation powers the X-rays. 

The high-mass X-ray binary LS~I~+61$^{\circ}$~303 (V615~Cas)  is known as a variable radio source (Gregory et al. 1979)
characterized by  nonthermal periodic radio outbursts (Gregory \& Taylor 1978). 
Its energetic outbursts are also visible  in  X-rays (Greiner \&  Rau 2001, Harrison et al. 2000), 
and GeV  wavelengths  (Abdo et al. 2009, Albert et al. 2008, Acciari et al. 2008).
The system has a relatively low X-ray luminosity for a 
high-mass X-ray binary,  but  is one of the 20 brightest  $\gamma$-ray sources known and detectable up to TeV energy range. 
It is one of only few such systems identified as a source of TeV gamma rays.
Indeed, it is currently considered to be a member of the new class
of gamma-ray-emitting binaries, that is binary systems whose luminosity
output has a dominating component at gamma-ray energies (see e.g. Paredes et al. 2013).


\lsi\ consists of  a massive  B0Ve star and a compact object. 
The nature of the compact object  remains a mystery even after four decades of observations over a 
wide range of  wavelengths.  
Most probably it is a magnetic neutron star, but it might  be a  magnetized  black hole (Punsly  1999) 
acting as a precessing microblazar (Massi, Ros \& Zimmermann 2012).  
The compact object  moving in an eccentric orbit interacts with the Be  circumstellar disk, producing strong orbital
modulation in the emission across the electromagnetic spectrum: radio flux (Gregory \& Taylor 1978; Taylor et al. 1992), 
X-ray (Paredes et al. 1997; Leahy 2001), optical V magnitude (Mendelson \& Mazeh 1994, Zaitseva \& Borisov 2003), 
TeV (Albert et al. 2006, 2008), and optical \Ha\  emission line (Zamanov et al. 1999, Grundstrom et al. 2007a). 
The radio morphology of the system also shows unique characteristics. The resolved extended structure changes position 
angle with a surprisingly large variation  (Massi et al. 2004).


The motion on the sky determined with  high precision astrometry implies that \lsi\ was ejected
from the stellar cluster IC~1805 by the kick imparted to the compact object in an asymmetric supernova explosion 
$\sim \! 1.7 \times 10^6$~yr ago (Mirabel, Rodrigues, \& Liu, 2004).

Additionally to the 26.5 d orbital modulation, a 4.4~yr periodic modulation of the phase and amplitude
of the radio  outbursts has been discovered (Paredes et al 1987; Gregory et al. 1989).
This superorbital modulation has been detected in \Ha\ (Zamanov et al.  1999). 
The first clues that the 4~yr modulation is also visible in X-rays have been given by Apparao (2001), and are now 
confirmed by Li et al. (2012)  using  the longest monitoring performed to date by the Rossi  X-ray Timing Explorer (RXTE).


In this paper we present  \Ha\  observations of LS~I~+61$^{\circ}$~303 
obtained during the past 15 years, 
and discuss  disk truncation, long-term variability,  orbital modulation, and connection with high-energy emission. 


 \begin{figure}   
  \vspace{7.5cm}   
  \includegraphics{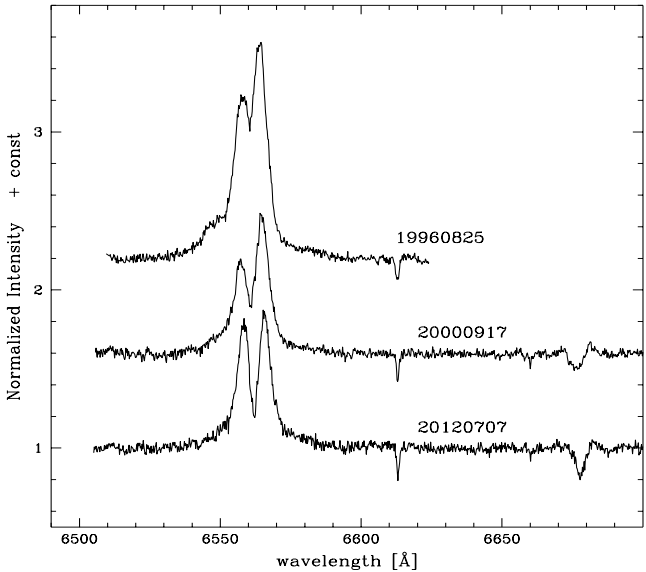}    
  \caption[]{Three examples of \Ha\  profiles. The  upper spectrum is  taken from the time of  maximal  EW=18.2 \AA\  (19960825).
   The middle spectrum has a  EW=13.1~\AA\ (2000917). The lower spectrum (20120707,  EW=10.8) is taken at a moment when 
   the star fulfilled  the shell criterion (see also Sect.~\ref{sect.shell}). }  
\label{fig.examp}      
\end{figure}	     

 \begin{figure}   
  \vspace{15.0cm}   
  \includegraphics{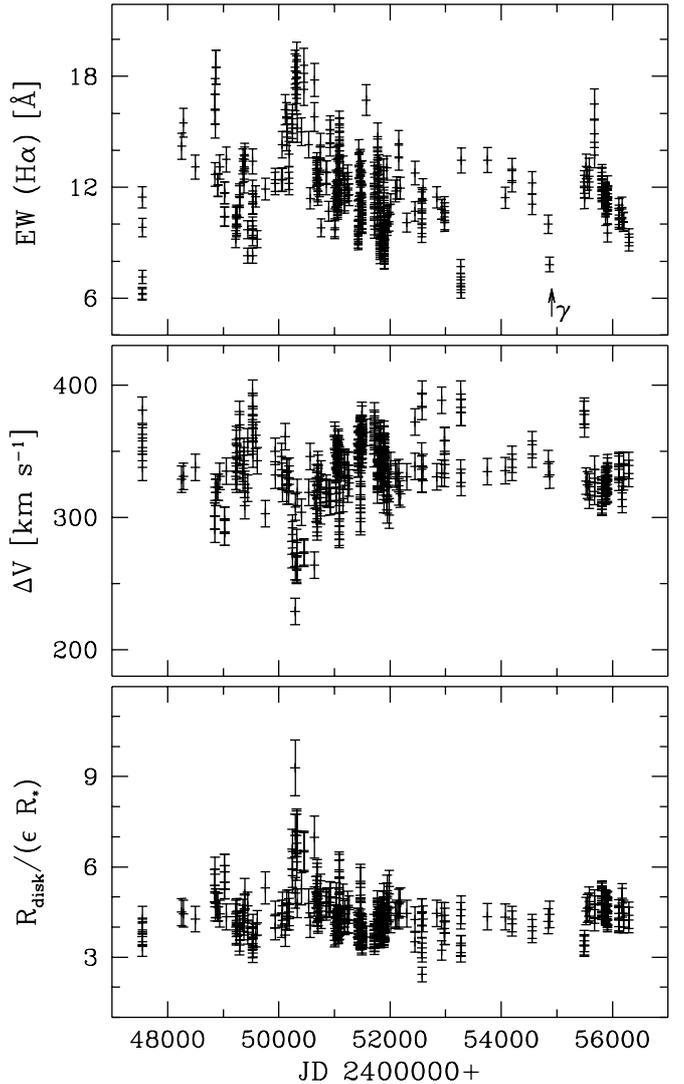}    
  \caption[]{Long-term variability of the equivalent width EW(\Ha ), the distance between the peaks
  of the \Ha\ emission line, and the calculated radius of the \Ha-emitting circumstellar disk. 
  The arrow in the upper panel corresponds to the time of the $\gamma$-ray flux transition 
  (see also Sect.\ref{sect.gamma}).   }  
\label{fig.long}      
\end{figure}	     

\section{Observations}

All  data reported here were obtained
by the 2.0m RCC  telescope of the National Astronomical Observatory Rozhen
located in the Rhodope mountain range, Bulgaria. 
The star \lsi\  was observed  between September  1998  and  January 2013 with the Coud\'e spectrograph of this telescope.
The spectra cover  200 \AA \ around H$\alpha$, with a resolution of  0.2 \AA\  pixel$^{-1}$.  

The spectra were reduced in the standard way including bias removal, 
flat-field correction, wavelength calibration, and correction for the Earth's motion. 
Pre-processings and measurements were performed using various routines provided by IRAF. 
The spectra obtained within each observational night were processed and measured independently. 

Table~\ref{tab.spec} lists the spectrum ID (the first six digits corresponding to the date - YYYYMMDD), 
JD of the start of the exposure, 
exposure time in minutes, 
signal-to-noise ratio (S/N) calculated for the continuum in the wavelength range  6620 - 6655 \AA.
On each spectrum we  measured the total equivalent width of the
H$\alpha$ emission line, hereafter $EW$, the heliocentric
radial velocities of the central dip $V_{r}(cd)$, blue  and red humps, $V_{r}(B)$ and $V_{r}(R)$ respectively, the ratio
between the equivalent widths of the blue and red humps, 
the intensity of the blue peak  $I_B$, 
the intensity of the red peak $I_R$, 
and the intensity of the central depression $I_{cd}$.  The intensities were measured after normalization to the local 
continuum, $I_C \equiv 1$.  The B/R ratio was calculated as  $B/R = I_B / I_R$ (this ratio 
is more often called V/R ratio, but most of the papers on \lsi\ use B/R).
The equivalent width of the blue hump, EW(B), was measured from the violet end 
to the central dip. The equivalent width of the red hump, EW(R), was measured 
from the central dip to the red end of the line, in this way:  $EW = EW(B)+EW(R)$.   
We also give radial velocity of the  HeI$\lambda 6678.151$ absorption line.  
The  radial velocities were measured by employing a Gaussian fit.  
The  errors depend mostly on the S/N, and we estimated them to be about 5\% for $EW$, about 10 \kms\ for
the radial velocities,  and 0.01 for the intensity.  

A few examples of our spectra are plotted in Fig.\ref{fig.examp}. The long-term 
variability of EW and $\Delta V$ are presented in Fig.\ref{fig.long}.

 \begin{figure}   
  \vspace{7.0cm}   
  \includegraphics{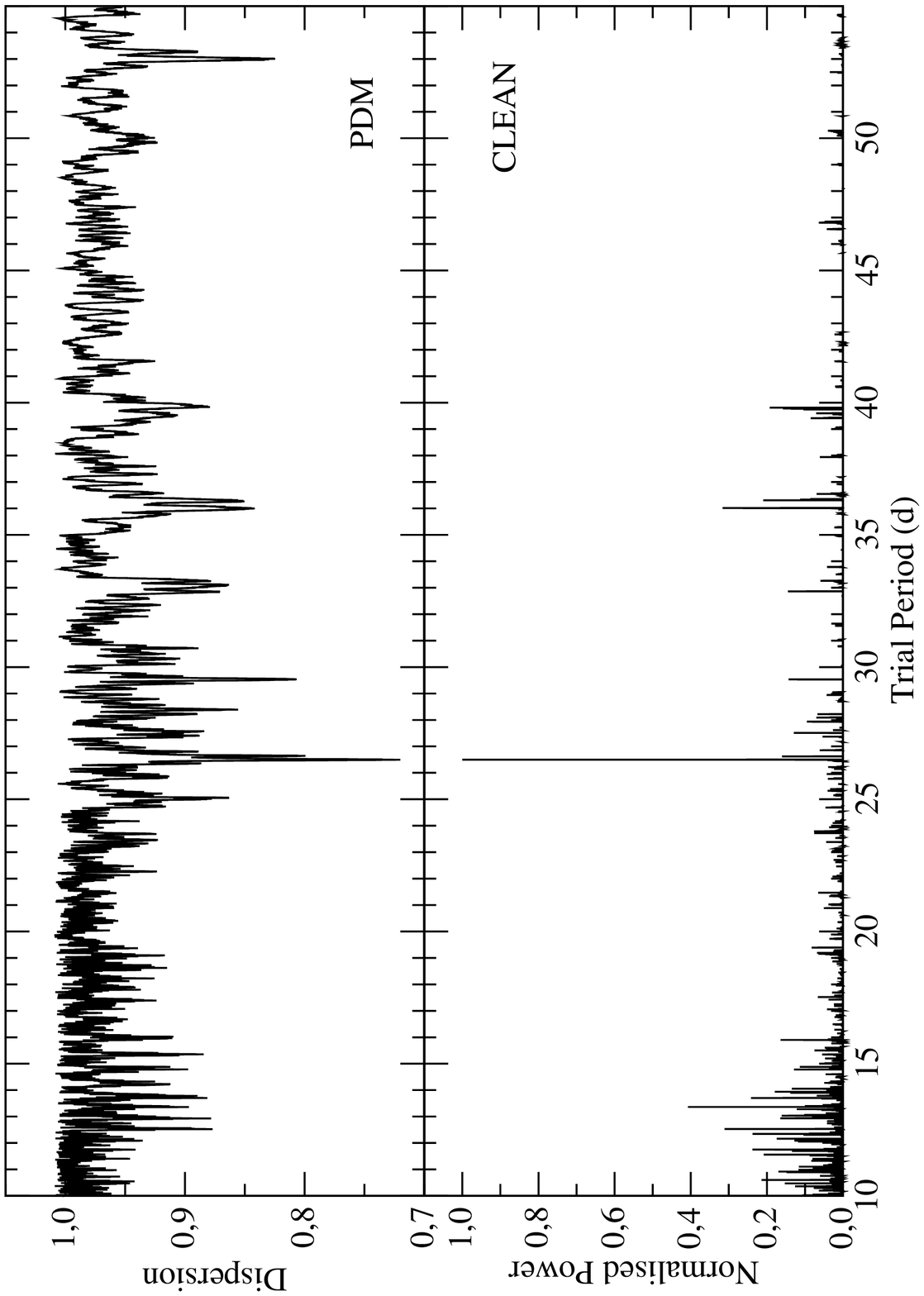}    
  \caption[]{Periodograms for the ratio  $EW_B/EW_R$  of \lsi. 
             They are computed using the PDM and CLEAN methods.
             The orbital period is detected in both cases as the most significant one. }  
  \label{fig.period1}      
  \vspace{7.5cm} 
  \includegraphics{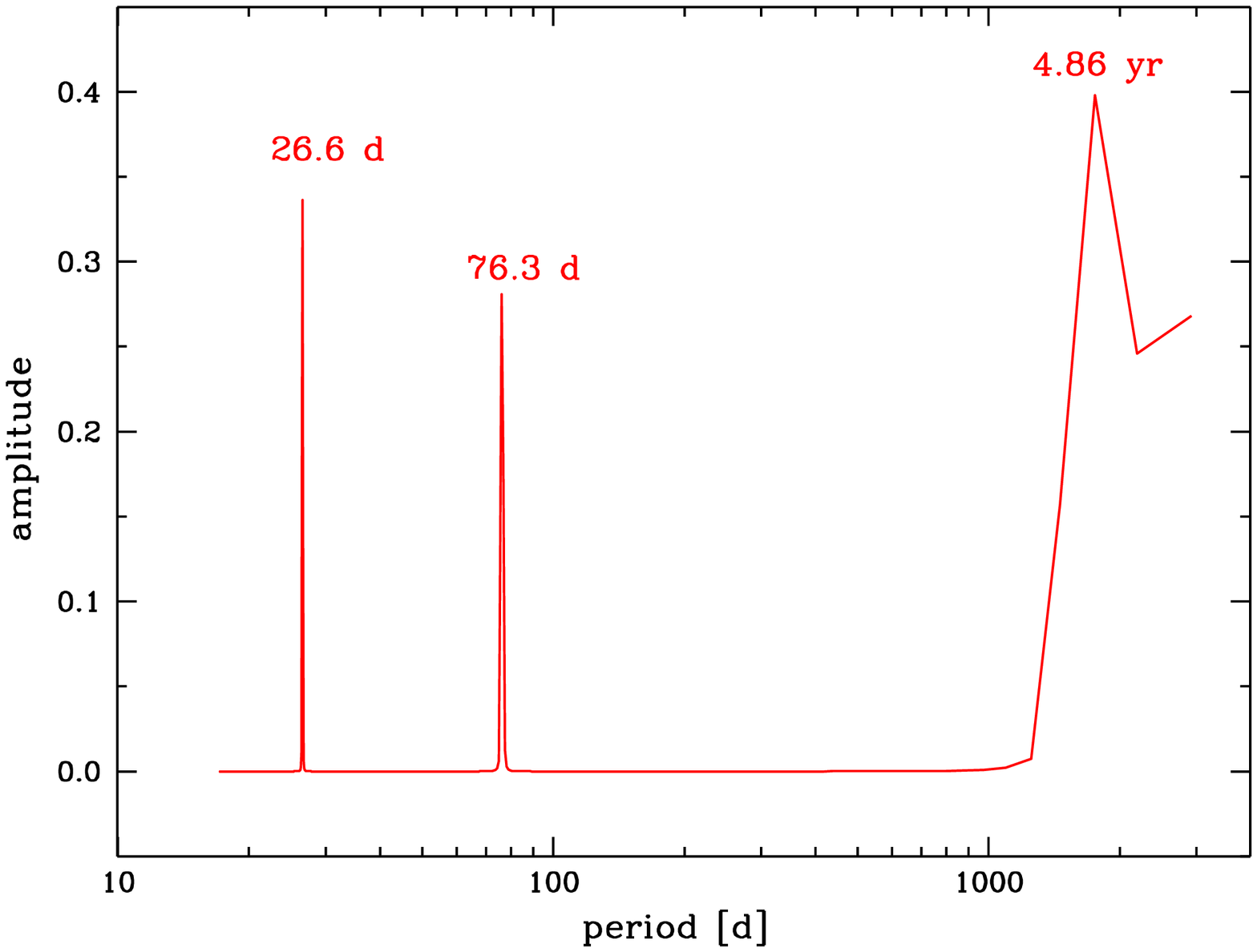}   
  \caption[]{Results of harmonic fit to the EW data, which clearly show  
             the orbital period and a 4.5 yr long-term modulation.}     
  \label{fig.Jaen}        
\end{figure}	     

 \begin{figure}   
  \vspace{5.4cm}   
  \includegraphics{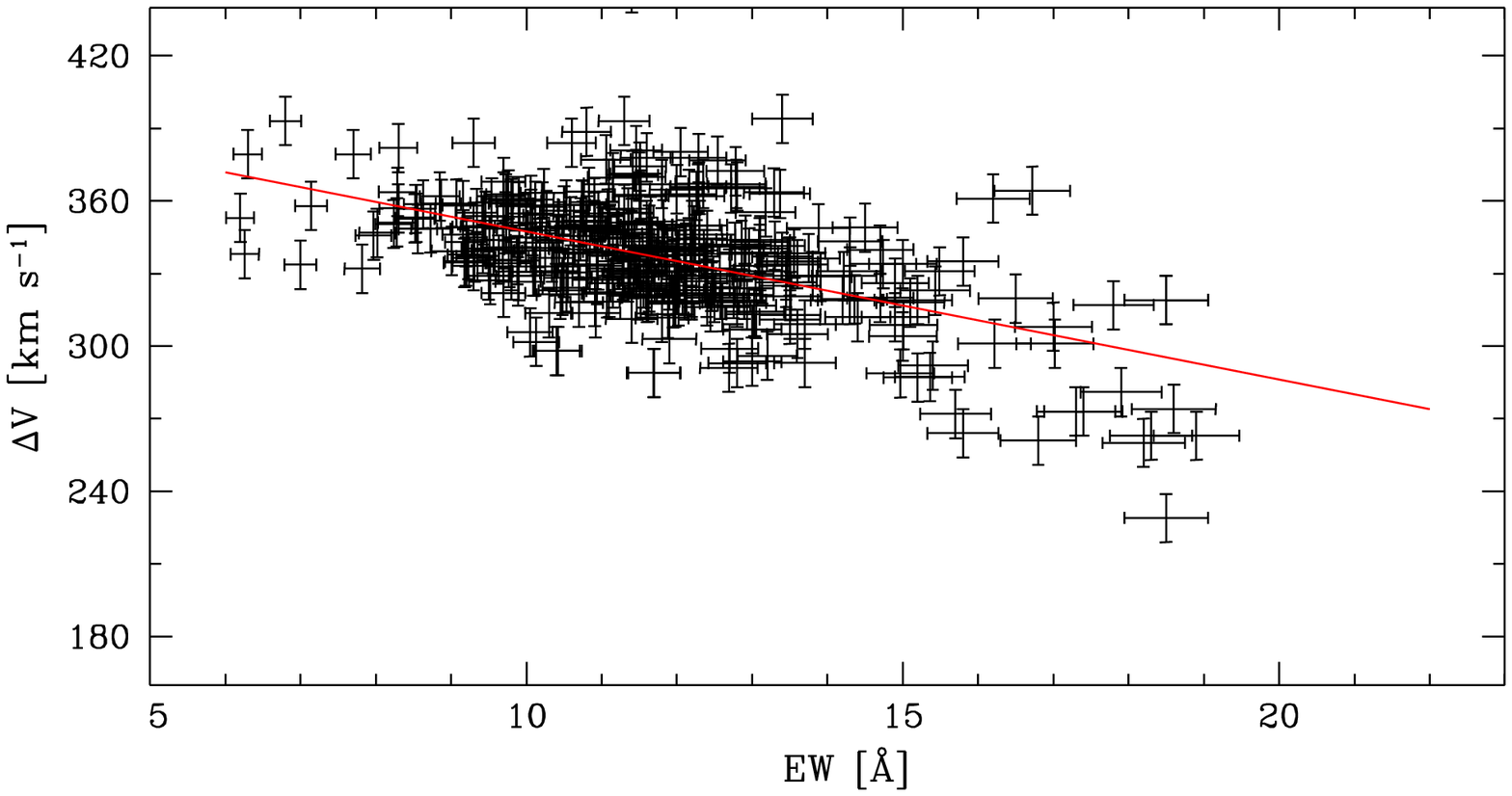}    
  \caption[]{$\Delta V$ versus EW (upper panel) and $R_{disk}$  vs.  EW (lower panel). The solid line  is the best
  linear fit (see Eq.\ref{eq.dV.corr}).
  }  
  \label{fig.EW.dV}      
  \vspace{5.4cm} 
  \includegraphics{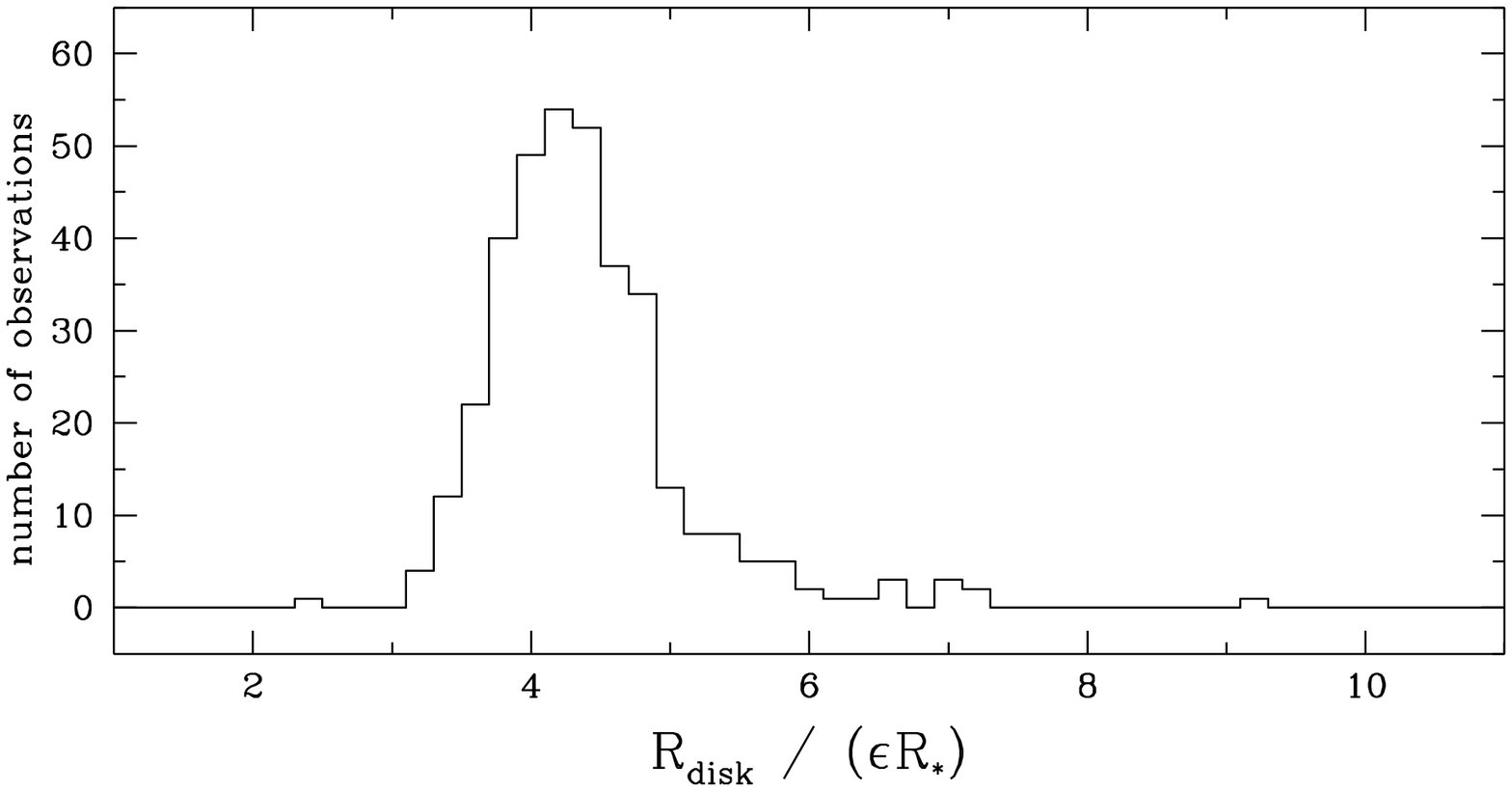}    
  \caption[]{Histogram of  $R_{disk}$  values. 
   The peak of the distribution  corresponds to  $R_{disk} \approx 26$~\rsun\ (see Sect.~\ref{Disk.size}). }
  \label{fig.hist}       
  \vspace{5.9cm} 
  \includegraphics{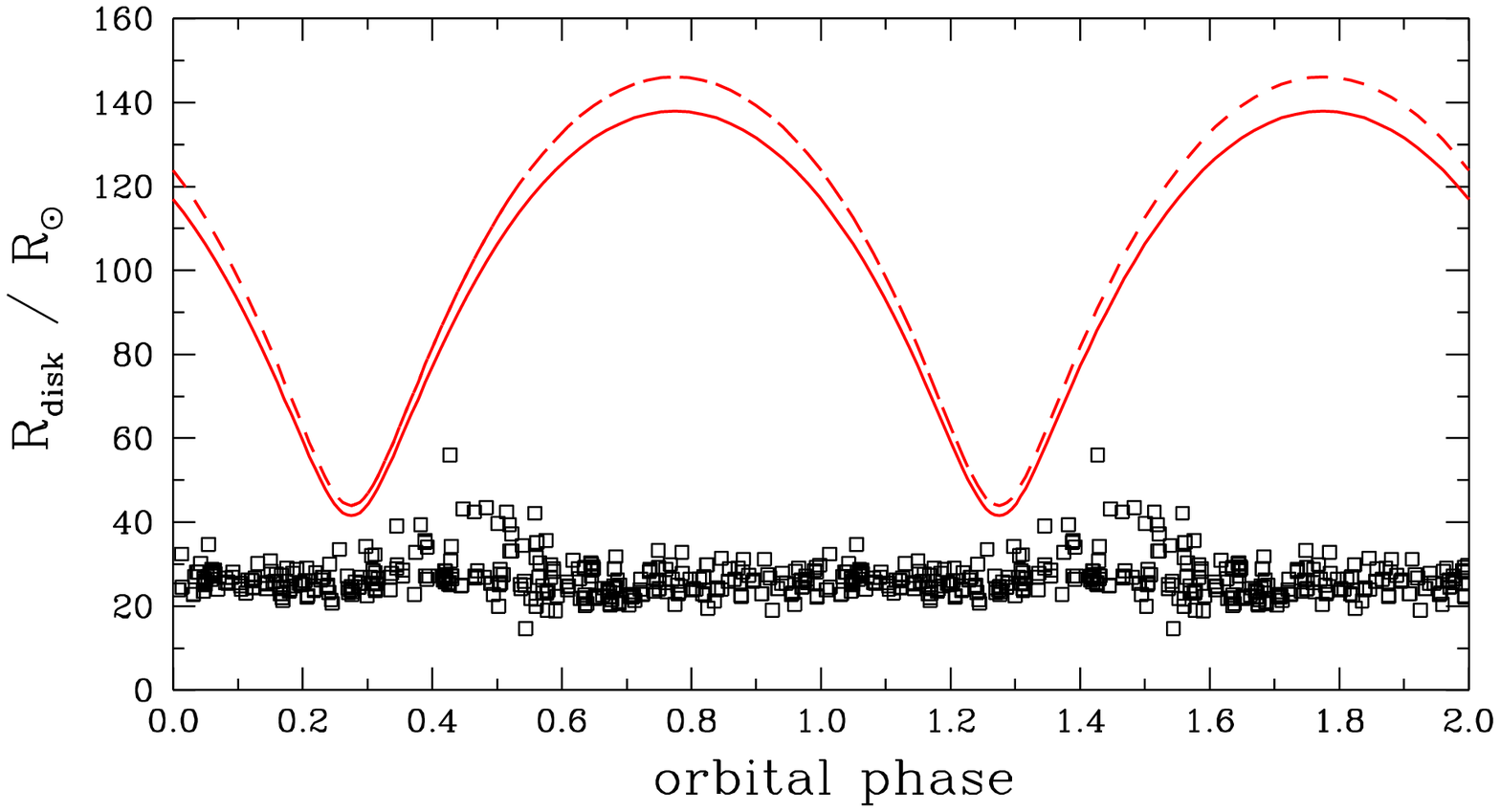}    
  \caption[]{Orbital variability of $R_{disk}$. The solid line is the distance between the components
  for a neutron star, the dashed line plots the distance to a black-hole companion.
  Interestingly, the orbital variability of $R_{disk}$  is not in phase with  the distance between 
  the Be star and the compact object.}
  \label{Rd.26d}      
\end{figure}	     

 \begin{figure}    
   \vspace{23.0cm}   
     \includegraphics{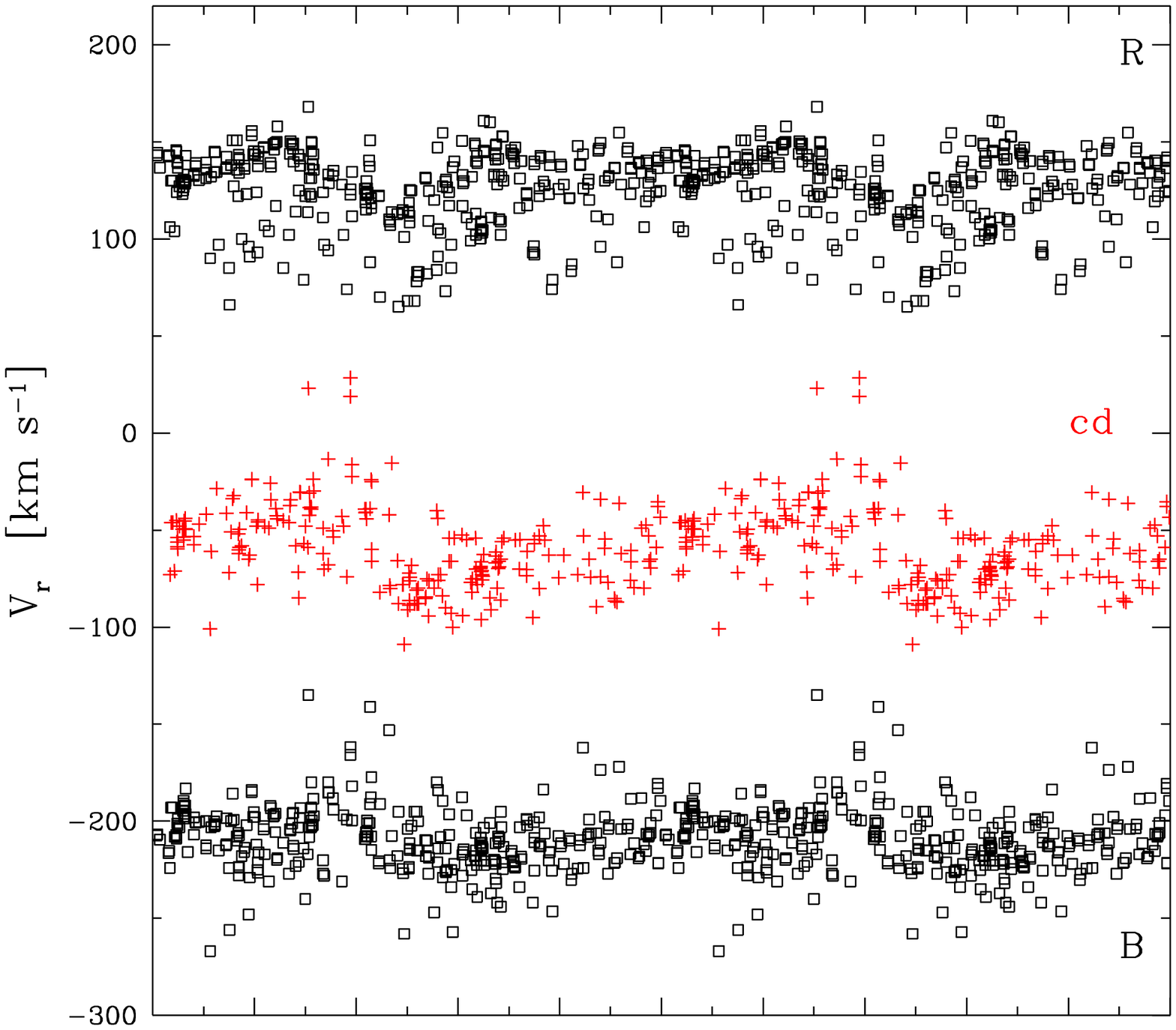}    
     \includegraphics{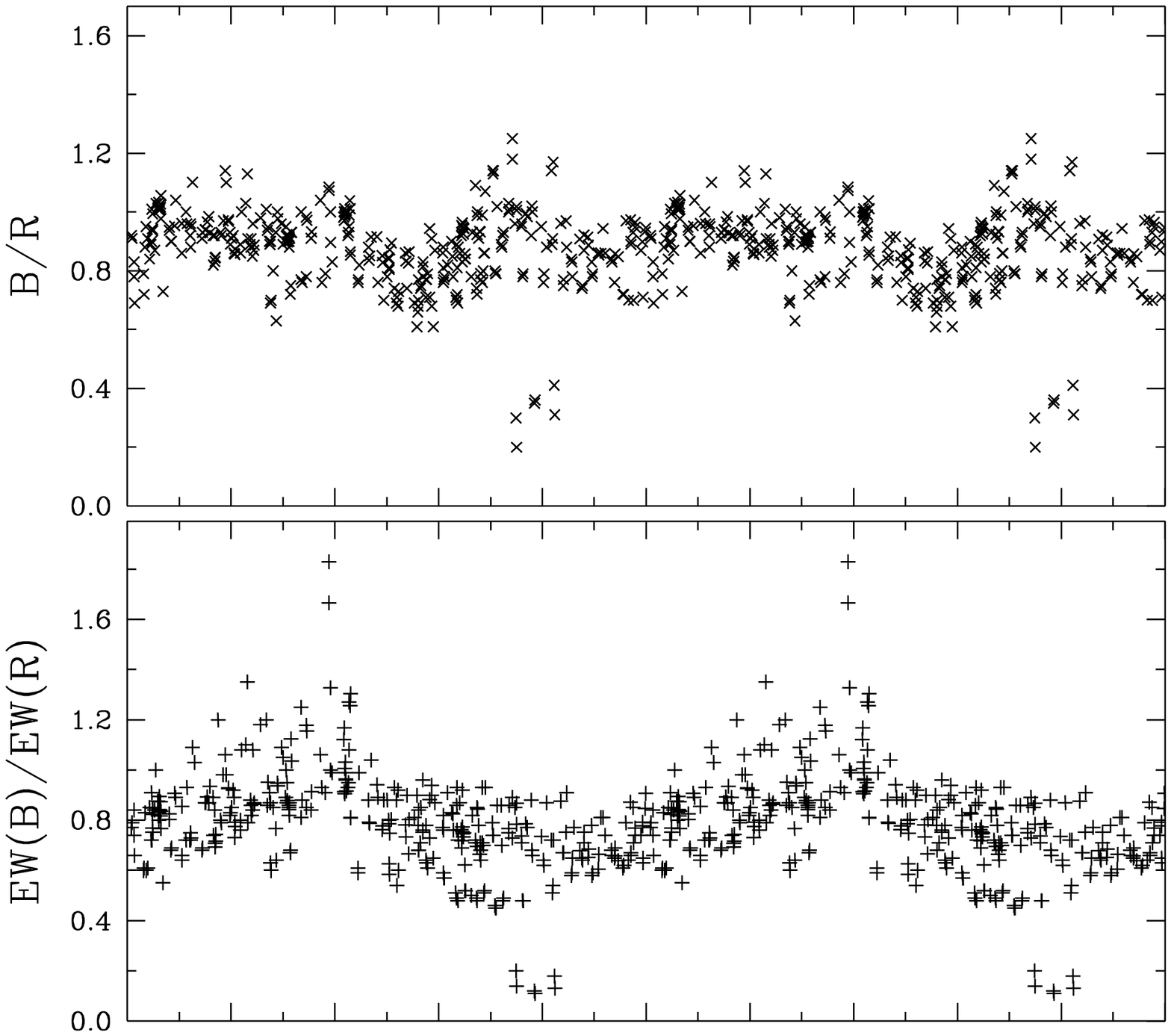}  
     \includegraphics{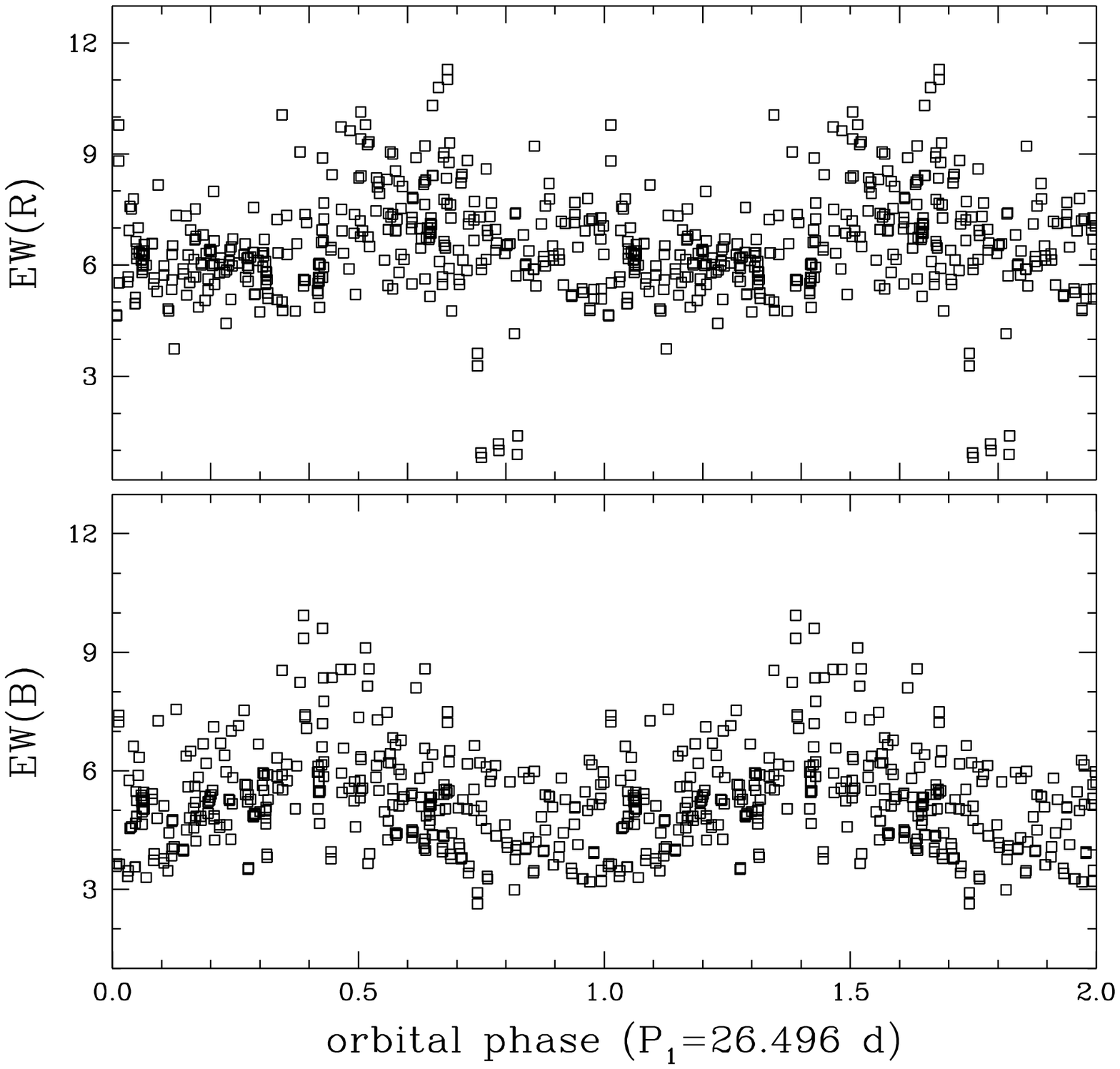}	
  \caption[]{ \Ha\  parameters  folded with  orbital period.  
      The following parameters are plotted  (from up to down):  radial velocity of the blue peak,  
      radial velocity of the red peak, B/R ratio, the ratio of the equivalent widths of the blue and red peak, 
      the equivalent width of the red peak,  and the equivalent width of the blue peak. 
      The radial velocities are  those after JD2448868. }
   \label{fig.BR}      
 \end{figure}	     

 \begin{figure}   
     \vspace{11.5cm}   
     \includegraphics{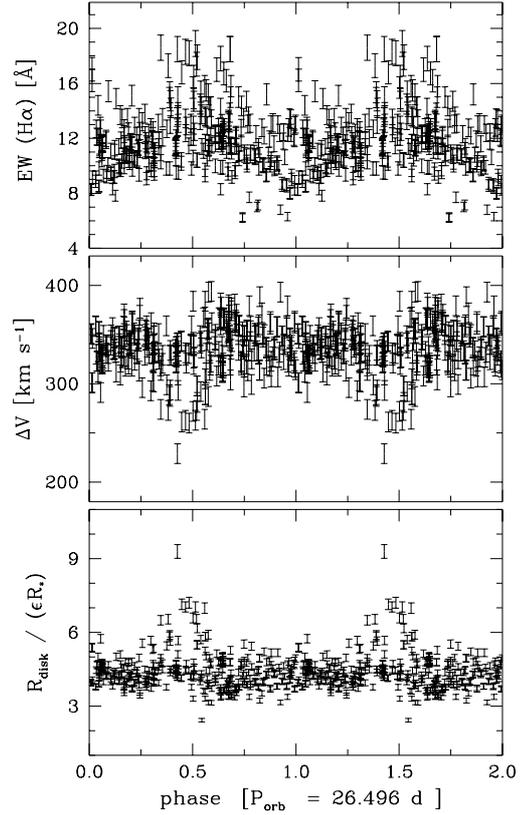}	
   \caption[]{ EW, $\Delta V$, and $R_{disk}$ folded with a period of 26.496~d.}	    
   \label{EW.orb}      
 \end{figure}      
%

 \begin{figure}   
   \vspace{8.0cm}   
   \includegraphics{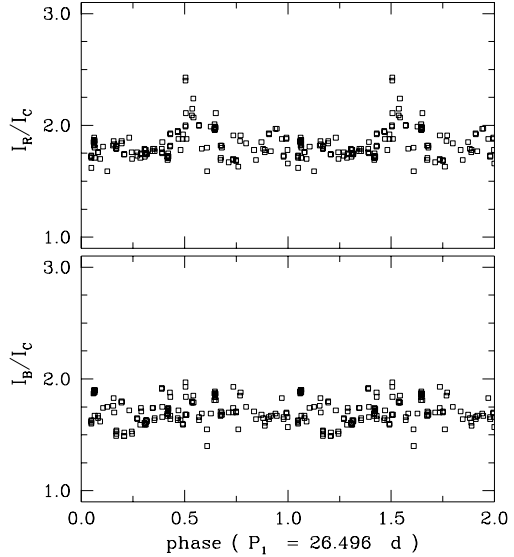}    
   \caption[]{Intensity of the red and blue peak  of \Ha\
        folded with  orbital period. 
	The orbital modulation is visible in $I_R$. These values are measured on our new spectra.}
\label{fig.IB.IR}      
\end{figure}	     

\section{Periodogram analysis of  the H$\alpha$ parameters}

\subsection{Detection of the orbital variability}

We conducted a period analysis for the different H$\alpha$ line parameters listed in Table 1, 
and also using data from McSwain et al. (2010), Grundstrom et al. (2007a), Liu \& Yan (2005),
Zamanov et al. (1999), Paredes et al. (1994), Steele et al. (1996).
The period-search methods were the phase dispersion minimization (PDM) (Stellingwerf 1978) 
and the CLEAN algorithm (Roberts, Lehar \& Dreher 1987). As a result, we confirm with a higher precision that the  
\Ha\ emission in LSI+61 303 displays variations with the orbital period.
The orbital modulation is better revealed in the ratio EW(B)/EW(R), EW(B), and  $V_{r}(cd)$.

In Fig.\ref{fig.period1} we present the PDM and CLEAN periodograms for  EW(B)/EW(R).
The most significant period detected in the range $10 - 60$~d corresponds to
26.498~d and 26.499~d for the PDM and CLEAN methods respectively. 
Using different  \Ha\ parameters, we obtain, an averaged value $26.502 \pm 0.007$~d, which is not an improvement over 
the radio orbital period. Thus, the H$\alpha$ emission line 
appears to match the same  orbital period of $26.4960  \pm 0.0028$~d 
obtained with Bayesian analysis of radio data (Gregory 2002).

\subsection{Long-term variability}

In addition to the orbital periodicity, another clock  is operating in \lsi, 
whose physical mechanism is not yet clear. The phase and amplitude of the radio outbursts
are known to exhibit a long-term 1600-day modulation (Paredes 1987; Gregory et al. 1989; Gregory 2002).

A visual inspection of Fig.\ref{fig.long} data  suggests that a  modulation of a few years may also be present.
To quantitatively address this superorbital-period question, all the PDM and CLEAN analyses were extended
up to 4000 d (i.e. half the full time-span of the data). The resulting periodograms (not shown here) suggested
possible frequency components around $\sim1800$ d. However, a precise determination of these long-term periodicities
was  difficulted due to the uneven sampling of the data and because only four superorbital cycles are covered. 
To obtain our best estimate, we applied a parametric harmonic modeling of the signal.
This can be achieved in a three-stage procedure. After detrending  the data,
the well-known Lomb-Scargle Fourier method was applied to estimate the frequencies and number of
components above the 99\% confidence level. Amplitudes and phases were then determined by means of a linear fit. The resulting parameters were
used as starting estimates for a nonlinear optimization. 
After building the component model, it was used to determine the definitive frequencies by means of      
traditional Fourier analysis  because now the model could be evenly sampled. 
As a final result, we obtained a refined estimate of the initially selected frequency components.

In our case, we maintained the three main frequencies resulting from the Lomb-Scargle algorithm. Residuals are within the 90\% critical limit,
thus indicating that they follow a Gaussian distribution, and therefore the model seems to be well-established.
Our best parametric harmonic modeling results are presented in the peridogram of Fig. 4. Here, the three main frequencies
are clearly detected finally above the 99.9\% confidence level. They correspond to periods of 26.6, 76.3,  and  1775~d. The first
value is obviously the orbital period, while the second one is close to, but not exactly coincident with, a harmonic of the former.
The third period revealed in Fig.\ref{fig.Jaen} corresponds to a 4.86~yr modulation, that is, very similar to the period originally
discovered at radio wavelengths. 

This is the first time that the LSI+61 303 superorbital period is blindly detected in the optical
spectroscopic properties of the source using periodogram techniques and thus confirms our early findings (Zamanov et al. 1999)
that \Ha\ is modulated with a superorbital period.


\section{Variability of the H$\alpha$ parameters}

\subsection{EW(\Ha)}

In our observations, the  EW(\Ha) of \lsi\  varies in the interval  7.8 - 18.9~\AA, 
with average $\overline {EW} = $  12.7~\AA, 
and standard deviation of the mean  $\sigma (EW) =$2.1 \AA. 

For the BeXRBs a correlation has been established between  the orbital period and 
the maximal EW (Reig, Fabregat \& Coe  1997). 
An updated version of this correlation  is given  by  Reig (2011).     
The maximum EW of \lsi\  observed until now (EW=18.9 \AA\ at JD~2450322) 
is  slightly higher than  the average behavior of the  Be/X-ray binaries. 

The linear regression  for the $\gamma$-ray binaries (see Eq.2 of  Casares et al. 2012a) predicts for 
\lsi\ a value for the  maximum EW$\approx $15.5~\AA. The observed value is again  slightly higher than the predicted
from the correlation. 

\subsection{B0V primary star}

For the primary we assumed  a B0V star of $M_1=12.5 \pm 2.5$~\msun\ 
(Hutchings \& Crampton 1981;  Casares et al. 2005). We also adopted 
$P_{orb}=26.496$~d and an eccentricity $e=0.537$ (Aragona et al. 2009). 
For the mass of the compact object, we assumed  $1.4$~\msun\ for a neutron star
and $4$~\msun\ for a black hole (see also Aragona et al. 2009). Using  Kepler's law\  \  
$4 \pi a^3 = G (M_1+M_2) P_{orb}^2$, we calculated the semimajor axis of the system.  

For the radius of the primary, using $M_1=12.5$~\msun\ and  Demircan \& Kahraman (1991) equations, 
we obtained   $R_1=4.3 $~\rsun\ for ZAMS,  $R_1=13.1 $~\rsun\ for TAMS.
The lifetime on the main sequence of a $12.5$~\msun\ star is $\; \sim \! 1.8 \times 10^7$~yr. 
On the other hand,  the cluster IC~1805 (where \lsi\ is probably born) is  young.  Its age is estimated 
to be  $\sim 3$~Myr (Massey, Johnson \& DeGioia-Eastwood 1995; Sung \& Lee 1995). 
This means that $R_1$ is expected to be closer to the ZAMS than to the TAMS value. 
For the radius of the primary, 
Grundstrom et al. (2007) adopted $R_1=6.7 \pm 0.9$~\rsun\ from the spectral classification and Harmanec (1988) tables.
This agrees with the above, and we adopted the same value.

Hanuschik (1989)  gives the relation  
 \begin{equation}
\log [FWHM(H\alpha) / (1.23\;  v\: \sin i + 70)] =  - 0.08 \log EW + 0.14, \\
 \end{equation}
where FWHM and \vsi\ are measured in \kms, EW is in [\AA] 
(see also   Reid \& Parker 2012). Using the spectra with  higher S/N ($S/N \ge 60$),
we measured a FWHM(\Ha) in the range $12.1 - 13.2$~\AA, and estimated  \vsi$=349 \pm 6$~\kms.
This agrees with the previous measurement \vsi=360 \kms\ by Hutchings \& Crampton (1981).

\subsection{Disk size}  
\label{Disk.size}

For rotationally dominated profiles the peak separation can be regarded as a measure of 
the outer radius of the \Ha\ emitting disk (Huang, 1972),  
 \begin{equation}
      \left( \frac{\Delta V}
                {2\,v\,\sin{i}} \;\right)
       = \;  \left( \frac {R_{disk}}{R_*}\;\right)^{-j} ,
  \label{Huang}
  \end{equation}
with $j=0.5$ for Keplerian rotation and $j=1$ for the conservation  of angular momentum.
We adopted $j=0.5$, because the kinematic evidence for Be stars points to a  velocity field dominated by Keplerian 
rotation, with little or no radial flow (Hummel \& Vrancken 2000; 
Hanuschik 2000; Porter \& Rivinius 2003 and references therein).
Eq.\ref{Huang} relies on the assumptions that (1) the Be star is rotating critically, 
and (2) that the line profile shape is dominated by kinematics, and radiative transfer does not play a role. 
The current view is that the Be stars rotate at values of 70\% - 80\% of the critical rate (Porter 1996; Chauville et al. 2001).
 Hummel \& Dachs (1992) showed that at higher optical depths the emission line peaks
 are shifted toward lower velocities, which means smaller peak separations. 
Hanuschik, Kozok \& Kaiser (1988) showed that the peak separation of H$\alpha$ line is smaller 
than the peak separation of the H$\beta$ line, and is probably smaller than the true kinematic value. 
Bearing in mind these effects, we used to calculate the disk radius
 \begin{equation}      
      R_{disk} = \epsilon \; R_* \frac{(2  v\,\sin{i} )^2}{\Delta V^2},
  \label{Huang2}
  \end{equation}
where $\epsilon$ is a dimensionless parameter, for which  we adopted $\epsilon = 0.9 \pm 0.1$.

Fig.\ref{fig.long} shows the long-term variability of the EW(\Ha), 
the distance between the red and blue peaks ($\Delta V$), 
and the size of the \Ha-emitting disk ($R_{disk}$). 
$\Delta V$ was calculated as  $\Delta V = V_r(R) - V_r(B)$, and  the disk size $R_d$  was calculated from Eq.\ref{Huang}.

In our observations  $\Delta V$ varies in the interval 137 - 394 \kms, with 
$\overline {\Delta V} = 329$  \kms, and $\sigma (\Delta V) = 30 $ \kms. 
This, following Eq.\ref{Huang2}, and assuming  $\epsilon = 0.9$, $R_*=6.7 $,
corresponds to   $R_{disk}$ in the interval  15 - 56 \rsun, 
90\% of the calculated values  are in the interval 21 - 34 \rsun, 
with  mean $\overline R_{disk} =$26.5~\rsun, median $<R_{disk}> \:  = 25.8$~\rsun, and $\sigma(R_{disk})=$4.4~\rsun.

\subsection{$\Delta V$  versus EW}

In Fig.\ref{fig.EW.dV} we plot $\Delta V$ versus the EW. 
The linear fit (of type y = a + bx) to the data points in Fig.\ref{fig.EW.dV}~(upper panel)  gives 
 \begin{equation}
   \Delta V  \;  [{\rm km\, s}^{-1}] = 408.3(\pm 6.0)  -  6.11(\pm 0.50) \; EW  \, [\AA] , 
    \label{eq.dV.corr}  
 \end{equation}
The errors of the coefficients are given in brackets.  The Spearman (rho) rank correlation gives  $\rho = -0.47$
(significance $10^{-9}$). The significance is  $<< 0.001$, indicating that  the correlation is
highly significant.  Eq.\ref{eq.dV.corr} was obtained on the base of 351 data pairs obtained during the past 21 years.
This correlation is much shallower when we use only the new data (Table~\ref{tab.spec}),
probably reflecting the fact that during the time of our new observations, the 4~yr modulation is obscured. 

The  negative correlation  between $\Delta V$ and EW(\Ha) expresses that the outer radius 
grows with increasing  $EW($H$\alpha)$ (e.g. Hanuschik, Kozok, \& Kaiser, 1988). 
A similar behavior is also visible in the \Ha\ observations of the Be/X-ray binaries LS~V$+44$~17 (Reig et al. 2005)
and  4U~$2206+54$  (Blay et al. 2006). 

 
\subsection{Disk truncation}

Haigh, Coe \& Fabregat (2004) and Coe et al. (2006) argued that the tendency for the disk emission fluxes to cluster at specified levels is
related to the presence of resonances between the disk gas and neutron star orbital periods 
that tend to truncate the disk at specific disk radii (Okazaki \& Negueruela 2001). These truncation
radii are given (see also Grundstrom et al. 2007b) by
  \begin{eqnarray} 
    {R_n^{3/2}} = \frac{(G \: M_1)^{1/2}}{2 \: \pi} \:  \frac{P_{orb}}{n}, 
  \label{eq.resona}
  \end{eqnarray} 
where $G$ is the gravitational constant and $n$ is the integer number of disk gas rotational periods  
per one orbital period. The important resonances are not 
only those with $n:1$, but can as well be $n:m$ in general, e.g, the 3:2 could be quite strong.

Okazaki \& Negueruela (2001) found that these limiting radii are defined by the closest approach of the companion
in the high-eccentricity systems and by resonances between the orbital period and the disk gas rotational periods in 
the low-eccentricity systems. \lsi\  has an orbital eccentricity that falls between these two cases, 
where the important resonance radii are similar in size to the periastron separation. 

We calculated for \lsi\  the  resonance  radii  as  
$R_8 =21.7$~\rsun, 
$R_7 =23.8$~\rsun,
$R_6 =26.3$~\rsun,
$R_5 =29.7$~\rsun.
$R_4 =34.5$~\rsun, and
$R_3 =41.8$~\rsun.
The peak of the  histogram (Fig.\ref{fig.hist}) corresponds to  $R_{disk} \approx 26$~\rsun, 
which corresponds to a 6:1 resonance. Bearing in mind the uncertainty of the adopted parameters, 
this peak is probably in the range $R_8 \le R_{disk} \le R_4$. 

The distance between the components at periastron is about 
$r(per) \approx 42$~\rsun\  for  $M_2=1.4$~\msun\  [$r(per) \approx 44$~\rsun\ for  $M_2=4$~\msun].
Using the Eggleton (1983) formula, we calculated  
a Roche lobe radius  $r_{RL} = 0.57$ in units of the orbital separation for $M_2=1.4$~\msun\  
and  $r_{RL} = 0.48$ for  $M_2=4$~\msun, respectively. 
This gives for a neutron star  $r_{RL} (per) = 24$~\rsun\  and $r_{RL} (per) = 78$~\rsun\ 
for the periastron and apastron, respectively
[for a black hole these values are  $r_{RL} (per) = 21$~\rsun\  and $r_{RL} (per) = 70$~\rsun].
For the circumstellar disk, it gives   $r_{RL}(per) \approx  R_{disk} < r(per)  < r_{RL}(ap)$.

 
For a similar system 4U~0115+63 ($P_{orb}=24.3$~d and $e=0.34$), the numerical simulations of 
Okazaki  et al. (2002) demonstraed that the surface density profile has breaks near the 5:1 resonance radius. 

For  the well-known X-ray pulsar  1A~0535+262 (HDE~245770, V725 Tau),  Grundstrom et al. (2007b)  
confirmed the expectations for resonance disk truncation. They found that 
the largest disk  radius  is similar to both the n=5  resonance radius and the mean Roche lobe radius at the time of periastron. 
The historical maxima of \Ha\  strength may imply that the disk  radius can occasionally grow to 
even larger dimensions.  

For \lsi\ the situation is similar. The average disk size is similar to 
both the $n=6 \pm 1$  resonance radius and the mean Roche lobe radius at the time of periastron. 
At times, it also grows to  higher values.  


\subsection{Orbital variability of the H$\alpha$ emission line}

The best  estimate of the orbital period of \lsi\  is  P$_{orb}= 26.4960 \pm 0.0028$ days derived 
from Bayesian analysis of the radio observations (Gregory 2002).
The zero phase is by convention JD$_0$=2,443,366.775, the date of the first radio detection of
the star (Gregory \& Taylor 1978). Using this ephemeris, we plot the orbital variability 
of different  \Ha\ parameters in Fig.~\ref{fig.BR}, Fig~\ref{fig.IB.IR} and \ref{EW.orb}.

According to the most recent radial velocity measurements, the orbit is elliptical ($e=0.537\pm0.034$),
and the  periastron passage is determined to occur around phase $\phi=0.275$, 
the apastron passage at $\phi=0.775$, the superior conjunction at $\phi=0.081$  (NS behind the B0V)
and inferior conjunction (the NS passes in front of the B0V star) 
at $\phi=0.313$ (Aragona et al. 2009). The eccentricity and relatively short orbital period provide significant changes in physical 
conditions along the orbit. 

The orbital modulation is  better visible in  the ratio  EW(B)/EW(R),  EW(B) and $V_r(cd)$.  
The highest-values  $EW(B)/EW(R) >1.0$ are reached in the orbital phase interval 0.13-0.43  when the neutron star 
is  on the left side (relatively to the line of sight) and moves toward the observer. The highest values of the $EW(B)$ are reached 
in the phase interval 0.3 - 0.6, 
the highest values of the total EW(\Ha) are also reached in the phase interval 0.3 - 0.6, 
where the minimum of $\Delta V$ and the maximum of  $R_{disk}$ occur as well.

The peak radio emission in each orbit occurs somewhere in range of phases extending from 0.4 to 0.9 (Gregory 2002). 
The onset of the outburst is  $\sim 0.1 - 0.15$ earlier (Zamanov \& Mart{\'{\i}}  2000). 
This means that the maxima of EW(\Ha)  and  $EW(B)$  occur more or less at the time of 
the start of the radio outbursts. 

One might expect that the disk size varies in phase with the distance between the Be star and the compact object.
However, this is not the case in \lsi. $R_{disk}$ 
is not in phase with  the distance between the Be star and the compact object. 
The peak of the disk size occurs in the interval 0.3-0.6,  that is, it preceedes the apastron,  which 
occurs at  $\phi=0.775$ (see Fig.\ref{Rd.26d} and Fig.~\ref{EW.orb}). This could be due to  size variation 
and/or some tidal noncircular motion in the disk (e.g.  Fig.~1 of Romero et al. 2007).


\section{Discussion} 

The circumstellar disks of the Be stars are inherently time variable and can develop
and disappear on timescales of years to decades (Underhill \& Doazan 1982; Hubert \& Floquet 1998).
Disk-loss episodes has been observed in the BeXRBs  X~Per (Roche et al. 1997) and 1A~0535+262 (Haigh et al. 2000). 
For \lsi\ there are \Ha\ observations for more then 20 yr,  during which period  no disk-loss episode 
has been observed.  The existence of  H$\alpha$ in emission 
throughout the past 24 years suggests that there is a permanent gas disk around the Be star.

\subsection{Orbital modulation of the EW(\Ha) in Be/$\gamma$-ray binaries}

Tarasov, Brocksopp \& Lyuty  (2003) demonstrated that the \Ha\ line-profiles 
of the supergiant X-ray binary Cyg~X-1 show a complex variability on different timescales,
controlled in particular by the orbital period and the focused wind model of the mass loss.

In our observations  the EW(\Ha) peaks at orbital phase $\phi \sim 0.4 - 0.7$ (see Fig. \ref{EW.orb}). 
The  detailed study of the \Ha\ emission of \lsi\ throughout an entire,
single orbit (McSwain et al. 2010) demonstrated  that there is a well-pronounced peak in 
EW(\Ha)  at orbital phases $0.5-0.7$. 
Bearing in mind that the periastron passage is at $\phi=0.275$, 
we conclude that the EW(\Ha) peaks at 0.2 - 0.4  phases past periastron.

Casares et al. (2012a) reported spectroscopic observations of the two  $\gamma$-ray binaries 
MWC~656 (AGL~J2241+4454) and  MWC 148 (HESS~J0632+057).
For MWC~656 (B3IV primary, $P_{orb} = 60.37$~d, $e\sim0.40$), they 
detected a maximum  of the EW(\Ha)  at  0.3 phases past periastron,
for MWC 148 (primary B0Ve, $P_{orb} =321$~d, $e \approx 0.83$), 
with the correct phasing (see Casares et al. 2012b), 
the maximum in the EW of the \Ha\ line occurs $\sim 0.5$ orbital phases past periastron.

Apparently it is a  common property of the  Be/$\gamma$-ray binaries that the maximum 
of the EW(\Ha ) occurs with a delay of  0.3 - 0.5 phases past periastron.

\subsection{Rotation of the components}

Be stars are fast rotators. They rotate at 70\% - 80\% of the critical rate with a rather
small intrinsic width of the distribution  (Porter \& Rivinius 2003 and references therein). 
The properties of the Be binaries with detected hot evolved companions
demonstrate that some fraction of Be stars were spun up through angular-momentum transfer by Roche-lobe 
overflow (Peters et al. 2013). 
This is probably  also the case for most of the Be/X-ray binaries. 
The rotation period of the  primary in \lsi\  (B0Ve star) is estimated to be $P_{rot} \sim 0.4 - 2.6$~d,
which means that it has a ratio $P_{rot} / P_{orb} \sim 0.05$, one of the higher values among the Be/X-ray binaries (Stoyanov \& Zamanov  2010).

The spin period of the compact object in \lsi is still unknown.
The deep searches in the radio- (McSwain et al. 2011, Ca{\~n}ellas et al. 2012) and X-ray band (Rea et al. 2010) did 
not detect  pulsed fraction so far.
The ejector-propeller model of \lsi\ predicts a spin period of the neutron star of 0.15-0.20~s (Zamanov 1995;
Zamanov, Marti \& Marziani,  2001).
One more prediction can be made using the Corbet diagram (Corbet 1986). The equation  
$\log P_{spin} = -1.011 +1.447 \log P_{orb}$ 
(where  $P_{spin}$ is in seconds, $P_{orb}$ is in days), which 
seems to be valid for wind-fed sources (X-ray pulsars in Be/X-ray binaries and white dwarfs in symbiotic stars), 
predicts  $P_{spin} \approx 11$~s.  


\subsection{Shell criterion and the Be star inclination}
\label{sect.shell}

The Be-shell stars are identified as normal Be stars viewed edge-on (Porter 1996;  Rivinius,  {\v S}tefl \& Baade 2006). 
Comparing  the FeII  and \Ha\  line profile,  
Hanuschik (1996) found that the shell stars are those with $I_p/I_{cd} (H\alpha) \ge 1.5$, 
where $I_p$ and  $I_{cd}$ are the mean peak intensity and the intensity of the central depression, respectively.

In our spectroscopic  observations of \lsi, $I_p/I_{cd} (H\alpha)$  varies in the range 1.16-1.70. Removing 
the four highest and  four lowest values,  we obtain a range of 1.20-1.53, a mean value of 1.40, and
a standard deviation of the mean 0.08. 
It falls in the intermediate cases according to Eq.3 of  Hanuschik (1996). 
Among our 130 measurements  there are about a dozen values with  $I_p/I_{cd} \ge 1.5$, and the Be star of \lsi\ fulfills 
the shell criterion  (e.g. 1998/10/04, 2012/07/07). 
The spectrum 20120707164 is plotted in the bottom panel of  Fig.\ref{fig.examp} and
it is visible that the central dip of \Ha\  intensifies and  the central absorption of HeI$\lambda$6678 deepens as well. 
The specific effect of the transition to the  Be-shell type  
is expected only at about an inclination angle $i \approx 70^\circ \pm 5^\circ$ (Hanuschik 1996). 
This suggests that the inclination of the primary star in \lsi\ to the line of sight is probably $i \sim 70^\circ$. 

One more clue for the inclination can be given using the full width at zero intensity (FWZI). 
For  an optically thin line this is connected with the inclination, 
$FWZI/2 \times \sin i = (GM_1/R_1) ^{1/2}$ (see Casares et al. 2012a and references therein). 
The FWZI of the \Ha\ line is  $\sim 60$~\AA, but it is not optically thin  and other mechanisms are broadening it. 
The FWZI of  HeI$\lambda 6678$   line is   19.0 $(\pm 10\%)$~\AA, which points again  to a high inclination  $i \sim 90^\circ$.

\subsection{Connection between \Ha\ and high-energy emission}

The nonthermal behavior of the source has been studied often, but is still poorly
understood. The nonthermal radio emission presents a well-defined periodicity
with strong flares that occur periodically near apastron passage, along
with an additional modulation on a 4.6 year timescale (Gregory
2002). The detection of extended structures in radio observations
originally identified LS~I~+61$^{\circ}$~303 as a potential
microquasar, with high-energy emission produced in jets driven by
accretion onto the compact object, presumably a black hole (Massi et  al. 2001). 
However, high-resolution VLBA observations indicate that the radio
structures are not persistent and can be more easily
explained by the interaction between a pulsar wind and the wind of the
stellar companion (Dhawan, Mioduszewski \& Rupen 2006, Albert et al. 2008), although
alternative interpretations are still possible (Romero et al. 2007, 
Massi and Zimmerman 2010). Many arguments in favor of 
LS~I~+61$^{\circ}$~303 as a nonaccreting pulsar system have been  summarized by Torres et al. (2010).
The high-energy process that causes the  very high energy emission could also produce neutrino flux. 
No evidence for periodic neutrino emission is found in  the IceCube Neutrino Observatory 
data  (Abbasi et al. 2012) until now.

\subsubsection{Connection between \Ha\ and $\gamma$-rays}
\label{sect.gamma}

The gamma-ray flux $(> 0.1 GeV)$  also displays  orbital modulation.  
Its maximum is at orbital phases 0.35 - 0.55  and the minimum at 0.9 - 1.0 (Hadasch et al. 2012). 
The maximum of the gamma-ray flux   coincides with 
the maxima  of the EW(\Ha), EW(B), and  $R_{disk}$ (see also Fig.\ref{EW.orb}). 
The   minimum of the gamma-ray flux corresponds to the minima of 
EW(\Ha) and EW(B). 

A flux change in gamma rays (Hadasch  et al. 2012)  
occurred in March 2009  (around JD 2454900). Before the transition, 
the modulation was clearly visible and was compatible with the previously   published phasogram, 
whereas afterwards, the amplitude of the modulation diminished. 
This transition in gamma-rays corresponds to a minimum value of the EW(\Ha ). 
On  February 6, 2009 we observed  an EW(\Ha)=7.8~\AA, which is considerably below the average value
(see also Fig.\ref{fig.long}). This is approximately  the minimum of  1600-day modulation of the EW(\Ha).

\subsubsection{Connection between \Ha\ and X-rays}

Torres et al. (2010) showed that the soft-X-ray emission from \lsi\ 
presents a periodic behavior at the orbital period, with a varying shape.
Profile variability is seen from orbit to orbit on a multi-year timescale. 
In their analysis the orbital peak of the soft X-rays (see Fig.2 of Torres et al. 2010)
is at phases  0.4 - 0.7. This  coincides well  with 
the  EW, EW(B) and $\Delta V$ peaks, which occur at the same orbital phases $\phi \sim 0.4 - 0.7$ (see Fig. \ref{EW.orb}). 

Simultaneous X-ray and radio observations show that periodic radio flares 
always lag the X-ray flare by $\Delta\phi \simeq 0.2$ (Chernyakova et al. 2012),  
a behavior predicted  by the ejector - propeller model. 
Additionally, intense X-ray flares have been observed (Smith et al. 2009 and Torres et al. 2010), during which
the flux increased by up to a factor of five and variability on a 
timescale of a few seconds was observed. But because of  the relatively large RXTE-PCA field-of-view ($\sim1^{\circ}$ FWHM),
it cannot be ruled out that these flares are caused by  an unrelated source in
the same field. If such  flares are generated 
in \lsi, they are probably the reason for the dramatic decline in the \Ha\ 
emission observed in October 1999, which may have been caused by a sudden ionization of the disk (see Grundstrom et al. 2007a).


\section{Conclusions}									      
		
We summarize the main results of our spectral observations of the Be/X-ray binary 
\lsi\  as follows:
\begin{enumerate}									      
   \item We measured various parameters of the \Ha\ emission line - equivalent widths, radial velocities, and intensities.
   \item  The periodogram analysis confirmed that the \Ha\ emission is modulated with 
          the orbital and superorbital periods. The values are practically identical to those
          detected in the radio observations. 
   \item  For  the past 20 years the radius of the circumstellar disk is similar to the 
          Roche lobe radius at  periastron. It is probably truncated by a resonance of $n=6 \pm 1$. 
   \item  The orbital maxima of the equivalent width of \Ha\ emission and  the radius of the circumstellar disk peak
          after the periastron and  coincide on average with the  X-ray and $\gamma$-ray maxima. 
   \item  For the B0V primary, we estimated a projected  rotational velocity  \vsi$=349\pm 8$~\kms. 
   Its inclination  to the line of sight  is probably about $70^\circ$.
\end{enumerate}
In future  it will be interesting 
to compare the \Ha\ variability of \lsi\  in detail with the variation in X-rays and $\gamma$-rays. 

\begin{acknowledgements}
We thank the anonymous referee for constructive comments.
 This work was supported in part by the OP "HRD", ESF and Bulgarian  Ministry of Education, Youth and Science
 under the contract BG051PO001-3.3.06-0047. 
 JM and PLLE  acknowledge support by grant AYA2010-21782-C03-03 from the Spanish Government, and
Consejer\'{\i}a de Econom\'{\i}a, Innovaci\'on y Ciencia of Junta de Andaluc\'{\i}a as research group FQM-322, 
as well as FEDER funds.
\end{acknowledgements}

\begin{table*}
\caption{H$\alpha$ observations of \lsi .}             
\label{tab.spec}      
\centering                          
\begin{tabular}{lrrlrrrrrrrrrr}        
\hline\hline                 
 &  &  \\                
spectrum ID  &JD -start& exp & S/N & EW   & $V_{r}(R)$& $V_{r}(B)$& V$_r$(cd) & EW(B/R)& $I_B$ & $I_R$& $I_{cd}$ & V$_r$HeI     \\ 	
yyyymmdd..   &         & min &     & \AA  & \kms     & \kms       & \kms       &	&	&      &          &  \kms        \\
\\	       
\hline           												       
\\
1998093003    & 51087.4054 & 20& 47&  14.97 & 122.9 & -165.9 & 18.7 &	1.666	&   1.92 &  1.79  &  1.49 &  ---   &  \\   
1998093004    & 51087.4196 & 20& 45&  15.36 & 125.5 & -161.8 & 28.3 &	1.828	&   1.91 &  1.76  &  1.47 &  ---   &  \\   
1998100157    & 51088.4664 & 20& 57&  15.01 & 121.1 & -187.7 &-24.0 &	1.255	&   1.88 &  1.81  &  1.39 &  -2.9  &  \\   
1998100158    & 51088.4922 & 20& 51&  13.70 & 115.9 & -177.2 &-25.0 &	1.305	&   1.84 &  1.82  &  1.38 &  -21.2 &  \\   
1998100159    & 51087.5063 & 20& 56&  13.00 & 111.7 & -181.9 &-22.5 &	1.327	&   1.75 &  1.75  &  1.38 &  -19.8 &  \\   
1998100417    & 51091.3185 & 20& 37&  13.30 & 131.7 & -209.8 &-85.3 &	0.782	&   1.80 &  2.09  &  1.29 &  -67.9 &  \\   
1998100418    & 51091.3327 & 20& 40&  14.50 & 131.1 & -217.9 &-84.4 &	0.734	&   1.79 &  2.15  &  1.32 &  -64.9 &  \\   
1998100438    & 51091.4910 & 20& 44&  14.30 & 124.6 & -218.7 &-76.1 &	0.803	&   1.79 &  2.07  &  1.27 &  -60.9 &  \\   
1998100439    & 51090.5051 & 20& 52&  14.70 & 125.2 & -214.7 &-74.4 &	0.749	&   1.84 &  2.11  &  1.29 &  -68.6 &  \\   
1998100970    & 51096.5846 & 20& 50&  11.66 & 123.6 & -199.2 &-73.6 &	0.750	&   1.70 &  1.70  &  1.26 &  -47.7 &  \\   
1998100971    & 51096.5988 & 20& 46&  12.75 & 122.9 & -198.7 &-70.2 &	0.766	&   1.74 &  1.69  &  1.27 &  -45.5 &  \\   
1998120914    & 51156.5455 & 20& 21&  13.03 & 124.7 & -189.5 &-43.4 &	0.847	&   1.66 &  1.78  &  1.29 &  ---   &  \\   
1998120915    & 51156.5596 & 20& 14&  12.00 & ---   &  ---   &---   &	0.908	&   1.57 &  1.66  &  ---  &  ---   &  \\   
1999010512    & 51184.3724 & 20& 30&  12.12 & 139.7 & -201.1 &-56.1 &	0.824	&   1.62 &  1.72  &  1.21 &  -44.2 &  \\   
1999010513    & 51184.3870 & 20& 37&  11.15 & 138.9 & -198.6 &-58.6 &	0.768	&   1.63 &  1.73  &  1.19 &  -34.9 &  \\   
1999010515    & 51184.4014 & 20& 37&  11.28 & 139.5 & -202.2 &-59.6 &	0.753	&   1.60 &  1.71  &  1.21 &  -48.7 &  \\   
1999010600    & 51185.2539 & 20& 42&  11.86 & 133.1 & -198.0 &-53.4 &	0.802	&   1.67 &  1.76  &  1.17 &  -31.9 &  \\   
1999010699    & 51185.2398 & 20& 53&  12.00 & 132.3 & -206.9 &-57.5 &	0.825	&   1.65 &  1.76  &  1.18 &  -38.9 &  \\   
1999030671    & 51244.2408 & 20& 44&  12.18 & 131.4 & -208.2 &-42.5 &	0.950	&   1.60 &  1.75  &  1.26 &  -36.6 &  \\   
1999030672    & 51244.2549 & 20& 31&  11.84 & 136.1 & -203.1 &-31.0 &	0.870	&   1.59 &  1.72  &  1.29 &  -20.3 &  \\   
1999030673    & 51244.2692 & 20& 42&  12.58 & 123.1 & -198.3 &-38.4 &	0.878	&   1.60 &  1.78  &  1.28 &  -35.4 &  \\   
1999032619    & 51264.2936 & 20& 42&  11.59 & 142.9 & -183.1 &-49.5 &	0.766	&   1.67 &  1.71  &  1.19 &  -27.8 &  \\   
1999091752    & 51439.4167 & 20& 48&  13.18 & 144.1 & -211.3 &-63.4 &	0.695	&   1.68 &  1.82  &  1.20 &  -46.8 &  \\   
1999091771    & 51439.5593 & 20& 52&  13.88 & 141.1 & -207.7 &-69.8 &	0.663	&   1.70 &  1.82  &  1.18 &  -53.0 &  \\   
1999091772    & 51439.5735 & 20& 57&  13.11 & 133.2 & -210.6 &-66.1 &	0.710	&   1.68 &  1.80  &  1.18 &  -48.9 &  \\   
1999091937    & 51441.4296 & 20& 46&  10.98 & 138.4 & -210.9 &-57.0 &	0.868	&   1.71 &  1.68  &  1.17 &  -32.9 &  \\   
1999091938    & 51441.4437 & 20& 45&  10.80 & 140.0 & -208.5 &-54.8 &	0.784	&   1.70 &  1.69  &  1.21 &  -38.2 &  \\   
2000012713    & 51571.2966 & 20& 67&  16.72 & 160.9 & -203.3 &-62.5 &	0.622	&   1.81 &  2.11  &  1.28 &  -66.9 &  \\   
2000062161    & 51717.4849 &  5& 23&  12.72 & 141.7 & -223.5 &-62.0 &	0.693	&   1.53 &  1.81  &  1.14 &  -9.6  &  \\   
2000062162    & 51717.4943 & 20& 41&  11.46 & 143.4 & -227.8 &-60.3 &	0.709	&   1.49 &  1.79  &  1.14 &  -2.7  &  \\   
2000062163    & 51717.5091 & 20& 40&   9.91 & 134.1 & -223.8 &-58.8 &	0.743	&   1.54 &  1.82  &  1.13 &  -5.6  &  \\   
2000062164    & 51717.5241 & 20& 53&  11.63 & 140.9 & -207.3 &-49.7 &	0.717	&   1.51 &  1.79  &  1.14 &  -28.5 &  \\   
2000062387    & 51718.4960 & 20& 40&  11.52 & 141.4 & -221.2 &-45.8 &	0.731	&   1.49 &  1.74  &  1.19 &  -11.5 &  \\   
2000062388    & 51718.5107 & 20& 56&  10.83 & 137.5 & -226.4 &-44.8 &	0.793	&   1.49 &  1.74  &  1.13 &  -29.3 &  \\   
2000062389    & 51718.5252 & 20& 45&   9.73 & 145.0 & -216.0 &---   &	0.775	&   1.52 &  1.74  &  1.13 &  -43.1 &  \\   
2000062393    & 51719.4816 & 20& 40&  11.15 & 150.0 & -219.8 &-42.6 &	0.864	&   1.53 &  1.76  &  1.19 &  -7.3  &  \\        	 
2000062395    & 51719.5108 & 20& 31&  12.54 & 157.9 & -218.9 &-44.2 &	0.869	&   1.51 &  1.70  &  1.19 &  ---   &  \\   
2000081776    & 51774.3832 & 16& 36&  11.15 & 143.5 & -188.5 &-23.8 &	1.124	&   1.63 &  1.75  &  1.21 &  -8.0  &  \\   
2000081777    & 51774.3956 & 20& 48&  10.37 & 135.6 & -197.3 &-30.0 &	1.035	&   1.62 &  1.74  &  1.23 &  -5.2  &  \\   
2000081819    & 51775.3947 & 20& 41&  13.51 & 132.5 & -188.1 &-50.4 &	0.840	&   1.63 &  1.79  &  1.34 &  -10.2 &  \\   
2000081820    & 51775.4073 & 15& 44&  12.03 & 135.1 & -194.0 &-53.5 &	0.914	&   1.65 &  1.77  &  1.30 &  -35.5 &  \\   
2000081985    & 51776.3989 & 20& 55&  14.74 & 134.5 & -199.5 &-16.4 &	1.000	&   1.66 &  1.85  &  1.29 &  -31.9 &  \\   
2000082038    & 51777.3920 & 20& 53&  13.91 & 120.9 & -207.9 &-59.9 &	0.811	&   1.64 &  1.92  &  1.28 &  -67.8 &  \\   
2000082039    & 51777.4082 & 20& 68&  13.10 & 118.4 & -214.7 &-66.1 &	0.808	&   1.67 &  1.93  &  1.29 &  -48.1 &  \\   
2000082105    & 51778.3539 & 20& 50&  13.45 & 109.0 & -220.1 &-78.6 &	0.792	&   1.63 &  1.95  &  1.30 &  -54.9 &  \\   
2000082106    & 51778.3681 & 20& 52&  12.38 & 107.2 & -224.2 &-80.2 &	0.788	&   1.65 &  1.94  &  1.30 &  -53.3 &  \\   
2000082206    & 51779.3545 & 20& 46&  12.09 & 118.7 & -223.5 &-72.1 &	0.786	&   1.68 &  2.00  &  1.26 &  -44.6 &  \\   
2000082207    & 51779.3687 & 20& 48&  12.96 & 114.2 & -224.5 &-76.1 &	0.800	&   1.68 &  1.99  &  1.29 &  -37.6 &  \\   
2000091723    & 51805.5042 & 20& 61&  12.10 & 115.0 & -224.6 &-78.0 &	0.763	&   1.60 &  1.88  &  1.27 &  -71.0 &  \\   
2000091724    & 51805.5261 & 20& 64&  13.10 & 113.5 & -226.8 &-77.9 &	0.777	&   1.59 &  1.92  &  1.29 &  -58.4 &  \\   
2000120547    & 51884.4345 & 20& 67&  12.90 & 113.7 & -207.5 &-15.5 &	1.040	&   1.72 &  1.88  &  1.45 &  -56.3 &  \\   
2000120669    & 51885.4853 & 20& 65&  12.48 & 124.8 & -202.2 &-68.2 &	0.803	&   1.68 &  1.88  &  1.37 &  -49.8 &  \\   
20001208200   & 51887.4365 & 20& 68&  10.92 & 117.1 & -196.5 &-54.0 &	0.882	&   1.69 &  1.79  &  1.22 &  -49.5 &  \\   
20010204161   & 51945.3274 & 20& 49&  13.05 & 129.6 & -183.7 &-47.7 &	0.785	&   1.85 &  1.86  &  1.33 &  -13.0 &  \\   
20010206208   & 51947.4100 & 20& 65&  10.05 & 143.7 & -162.1 &-30.6 &	0.752	&   1.64 &  1.69  &  1.22 &  -50.0 &  \\   
20010207226   & 51948.2957 & 20& 61&  10.48 & 149.5 & -173.5 &-34.2 &	0.753	&   1.65 &  1.79  &  1.18 &  -28.1 &  \\   
& \\
\hline                                   
\end{tabular}
\end{table*}
%
\addtocounter{table}{-1} 
\begin{table*}
\caption{Continued.}             
\label{table1cont}                     
\centering                          
\begin{tabular}{lrrlrlccccccrccc}   
\hline                        
&  \\
20010208261   & 51949.2902 & 20& 71&   9.89 & 154.8 & -171.9 &-36.3 &	0.811	&   1.67 &  1.77  &  1.13 &   -16.4 & \\   
20010317356   & 51986.2288 & 20& 51&  10.13 & 121.7 & -180.0 &-39.3 &	0.851	&   1.63 &  1.78  &  1.15 &   -38.4 & \\   
20010317357   & 51986.2458 & 20& 42&  10.53 & 131.3 & -193.0 &-38.2 &	0.843	&   1.59 &  1.74  &  1.18 &   -34.0 & \\   
2001040730    & 52007.2424 & 20& 40&  11.09 & 139.4 & -199.8 &-41.8 &	0.855	&   1.74 &  1.81  &  1.19 &   -33.8 & \\   
2001070958    & 52100.5735 & 20& 53&  11.79 & 107.7 & -225.2 &-82.1 &	0.826	&   1.69 &  1.99  &  1.28 &   -58.5 & \\   
2001072790    & 52118.5316 & 20& 43&  11.97 & 113.8 & -217.0 &-59.0 &	0.890	&   1.61 &  1.79  &  1.23 &   -56.0 & \\   
20010903161   & 52156.4462 & 20& 44&  14.36 & 117.3 & -202.1 &-61.0 &	0.860	&   1.93 &  1.91  &  1.31 &   -50.3 & \\   
20010904214   & 52157.3702 & 20& 50&  13.60 & 127.7 & -206.2 &-55.1 &	0.770	&   1.88 &  1.91  &  1.28 &   -59.1 & \\   
2001100323    & 52186.5529 & 20& 61&  11.95 & 111.6 & -206.1 &-89.5 &	0.679	&   1.68 &  1.85  &  1.28 &   -83.5 & \\   
2002012348    & 52298.3445 & 20& 65&  10.09 & 130.3 & -200.5 &-47.0 &	0.908	&   1.62 &  1.70  &  1.21 &   -52.9 & \\   
2002062229    & 52448.5433 & 20& 44&  10.69 & 126.0 & -212.6 &-80.1 &	0.738	&   1.55 &  1.63  &  1.19 &   -29.1 & \\   
2002062418    & 52450.5460 & 20& 39&  12.78 & 147.9 & -224.4 &-73.0 &	0.876	&   1.71 &  1.78  &  1.17 &   -26.5 & \\   
2002102031    & 52568.4718 & 20& 58&  11.20 & 143.3 & -203.2 &-71.7 &	0.766	&   1.59 &  1.71  &  1.25 &   -6.4  & \\   
2002111224    & 52591.4611 & 20& 58&  11.87 & 137.9 & -199.8 &-51.0 &	0.892	&   1.76 &  1.82  &  1.28 &   -18.0 & \\   
2003071728    & 52838.5832 & 20& 65&  11.46 & 112.8 & -217.8 &-65.8 &	0.942	&   1.63 &  1.78  &  1.29 &   -63.7 & \\   
20031205111   &	52979.461  & 20& 73&  10.95 & 125.7 & -207.6 &-74.5 &   0.735   &   1.75 &  1.84  &  1.21 &   -48.7 & \\
20031208143   &	52982.393  & 20& 34&  10.12 & 136.7 & -221.4 &-85.2 &   0.606   &   1.66 &  1.93  &  1.23 &   -59.3 & \\
20031208144   &	52982.408  & 20& 38&  10.21 & 136.4 & -221.3 &-86.6 &   0.664   &   1.64 &  1.92  &  1.18 &   -57.0 & \\
20031209181   &	52983.399  & 20& 43&  10.62 & 128.8 & -211.3 &-79.6 &   0.638   &   1.67 &  1.97  &  1.22 &   -72.1 & \\
20041001      & 53280.5038 & 20& 55&  13.45 & 127.1 & -199.2 &-32.4 &	0.935	&   1.70 &  1.83  &  1.20 &   -33.5 & \\   
20060116      & 53752.4002 & 19& 33&  13.47 & 119.4 & -215.4 &-47.5 &	0.871	&   1.83 &  1.88  &  1.31 &   ---   & \\   
20061202      & 54072.4667 & 20& 28&  11.43 & 136.6 & -198.7 &-53.8 &	0.849	&   1.61 &  1.62  &  1.15 &   ---   & \\   
2007040116    & 54192.2496 & 20& 19&  15.84 &  ---  & ---    &---   &	0.760	&   1.63 &  2.00  &  1.18 &   ---   & \\   
20070401      & 54192.2496 & 20& 19&  12.91 & 154.6 & -189.4 &---   &	0.752	&   1.66 &  2.00  &  1.28 &   ---   & \\   
2007040232    & 54193.2420 & 20& 31&  12.36 & 150.4 & -187.6 &-52.4 &	0.772	&   1.55 &  1.80  &  1.20 &   ---   & \\   
2007040236    & 54193.2700 & 20& 15&  12.58 & ---   & ---    &---   &	0.760	&   1.40 &  1.59  &  1.11 &   ---   & \\   
2008012541    & 54491.4239 & 20& 31&   9.64 & ---   & -206.2 &-74.2 &	0.772	&   ---  &  ---   &  1.07 &   -74.1 & \\   
2008032959    & 54555.2378 & 20& 39&  12.23 & 143.9 & -204.1 &-37.4 &	0.860	&   1.65 &  1.76  &  1.27 &   ---   & \\   
2008032965    & 54555.2603 & 20& 47&  11.08 & 150.3 & -204.6 &-34.5 &	0.952	&   1.64 &  1.73  &  1.28 &   ---   & \\   
2009011162a   & 54843.3744 & 20& 39&  10.00 & 142.3 & -198.4 &-41.2 &    ---	&   ---  &  ---   &  1.25 &   -45.2 & \\   
20090206104   & 54869.3824 & 20& 30&   7.82 & 134.6 & -197.4 &-28.5 &  1.090	&   1.75 &  1.59  &  0.95 &   -27.2 & \\   
2010102374    & 55493.3172 & 20& 57&  12.29 & 148.2 & -229.7 &-66.9 &  0.850	&   1.67 &  1.69  &  1.24 &   ---   & \\   
2010102377    & 55493.3374 & 20& 76&  12.05 & 148.6 & -231.7 &-61.3 &  0.845	&   1.71 &  1.71  &  1.26 &   -45.9 & \\   
20101024129   & 55494.3266 & 20& 57&  11.41 & 147.0 & -223.8 &-55.1 &  0.859	&   1.70 &  1.67  &  1.27 &   -17.5 & \\   
2010112406    & 55525.4759 &  4& 13&  13.56 & ---   &  ---   &-59.5 &  0.652	&   1.61 &  1.76  &  1.23 &   ---   & \\   
2010112407    & 55525.4847 & 20& 35&  13.13 & 129.9 & -197.7 &-54.5 &  0.686	&   1.58 &  1.78  &  1.16 &   ---   & \\   
2010122228    & 55553.3332 & 20& 53&  12.76 & 136.7 &  ---   &-66.2 &  0.658	&   0.00 &  0.00  &  1.26 &   ---   & \\   
2010122229    & 55553.3474 & 20& 47&  12.46 & 136.0 & -188.5 &-60.6 &  0.690	&   1.69 &  1.97  &  1.25 &   -55.5 & \\   
2011011923    & 55581.2366 & 20& 58&  12.00 & 140.2 & -182.9 &-38.0 &  0.791	&   1.69 &  1.88  &  1.23 &   -38.3 & \\   
2011011924    & 55581.2515 & 20& 62&  12.93 & 136.3 & -180.6 &-35.6 &  0.777	&   1.70 &  1.89  &  1.22 &   -34.5 & \\   
20110422107   & 55674.2894 & 20& 52&  16.50 & 108.3 & -211.4 &-88.6 &  0.627	&   1.93 &  2.40  &  1.58 &   -55.4 & \\   
20110422108   & 55674.3063 & 20& 53&  14.90 & 112.0 & -214.1 &-86.3 &  0.584	&   1.97 &  2.43  &  1.59 &   -68.3 & \\   
20110423145   & 55675.2838 & 20& 34&  13.75 & 109.3 & -227.0 &-94.2 &  0.667	&   1.85 &  2.24  &  1.44 &   -79.5 & \\   
20110905123   & 55810.4711 & 20& 57&  11.80 & 100.3 & -211.4 &-69.8 &  0.763	&   1.81 &  1.97  &  1.33 &   -39.5 & \\   
20110905124   & 55810.4861 & 20& 56&  11.40 & 100.3 & -211.0 &-69.5 &  0.689	&   1.87 &  2.01  &  1.31 &   -39.9 & \\   
20110905125   & 55810.5003 & 20& 57&  11.88 & 104.9 & -208.3 &-68.6 &  0.718	&   1.87 &  1.96  &  1.29 &   -36.1 & \\   
20110905126   & 55810.5144 & 20& 52&  12.20 & 108.6 & -212.3 &-66.2 &  0.774	&   1.84 &  1.98  &  1.35 &   -44.6 & \\   
20110905127   & 55810.5286 & 20& 53&  12.02 & 104.2 & -214.4 &-71.3 &  0.754	&   1.89 &  1.96  &  1.32 &   -49.7 & \\   
20110905128   & 55810.5427 & 20& 67&  12.45 & 105.0 & -211.1 &-70.9 &  0.746	&   1.86 &  1.96  &  1.33 &   -47.3 & \\   
20110905130   & 55810.5583 & 20& 64&  12.60 & 103.2 & -213.0 &-73.1 &  0.733	&   1.86 &  1.97  &  1.33 &   -42.9 & \\   
20110905131   & 55810.5724 & 20& 60&  11.98 & 104.5 & -213.6 &-75.7 &  0.749	&   1.84 &  1.99  &  1.32 &   -47.6 & \\   
20110905132   & 55810.5869 & 20& 69&  12.40 & 104.4 & -214.1 &-73.8 &  0.751	&   1.87 &  1.98  &  1.32 &   -56.8 & \\   
20111108059   & 55874.4505 & 20& 57&  11.52 & 123.0 &  ---   &-49.6 &  0.875    &   1.87 &  1.85  &  1.35 &   -16.5 & \\   
20111108060   & 55874.4661 & 20& 55&  11.60 & 124.9 & -194.6 & ---  &  0.818    &   1.89 &  1.86  &  1.36 &   -28.8 & \\   
20111108061   & 55874.4956 & 20& 65&  10.94 & 130.5 & -189.5 &-53.5 &  0.831    &   1.88 &  1.82  &  1.35 &    ---  & \\   
20111108062   & 55874.5106 & 20& 60&  10.45 & 126.6 & -190.9 &-55.1 &  0.801    &   1.89 &  1.87  &  1.37 &   -37.8 & \\   
20111108063   & 55874.5279 & 20& 56&  11.63 & 129.7 & -194.5 &-45.2 &  0.842	&   1.90 &  1.89  &  1.35 &   -19.9 & \\   
20111108064   & 55874.5428 & 20& 49&  11.15 & 128.5 & -197.5 &-49.2 &  0.894	&   1.89 &  1.85  &  1.38 &    ---  & \\   
20111108065   & 55874.5579 & 20& 56&  11.07 & 129.2 & -192.5 &-45.4 &  0.840	&   1.88 &  1.86  &  1.35 &   -25.4 & \\   
20111108066   & 55874.5725 & 20& 56&  11.62 & 128.3 & -191.9 &-43.9 &  0.818	&   1.90 &  1.87  &  1.36 &   -27.4 & \\   
20111108067   & 55874.5870 & 20& 52&  11.71 & 130.1 & -198.6 &-49.4 &  0.873	&   1.88 &  1.83  &  1.35 &   -39.2 & \\   
& \\
\hline                                   
\end{tabular}
\end{table*}
%
\addtocounter{table}{-1} 
\begin{table*}
\caption{Continued.}             
\label{table1cont}                     
\centering                          
\begin{tabular}{lrrlrlccccccrccc}   
\hline                        
&  \\
20111108068   & 55874.6016 & 20& 50&  11.56 & 129.6 & -196.3 &-49.7 &  0.837	&   1.90 &  1.80  &  1.32 & -38.1 &   \\   
20111214176   & 55910.4263 & 20& 39&  11.28 & 125.8 & -209.2 &-39.2 &  1.122	&   1.69 &  1.72  &  1.41 & -11.7 &   \\   
20111214179   & 55910.4459 & 20& 49&  11.35 & 120.5 & -207.9 &-41.0 &  1.167	&   1.69 &  1.69  &  1.43 & -1.6  &   \\   
20111214180   & 55910.4600 & 20& 31&  10.55 & 118.9 & -209.3 &---   &  0.914	&   1.67 &  1.72  &  1.41 & -0.2  &   \\   
20111214181   & 55910.4742 & 20& 52&  11.38 & 123.1 & -207.7 &-40.7 &  1.006	&   1.69 &  1.71  &  1.46 & -17.6 &   \\   
20111214182   & 55910.4883 & 20& 50&  11.47 & 115.2 & -205.8 &-39.3 &  0.905	&   1.73 &  1.71  &  1.44 & -0.8  &   \\   
20111214183   & 55910.5025 & 20& 52&  12.02 & 129.7 & -208.0 &-44.2 &  0.989	&   1.74 &  1.75  &  1.44 & ---   &   \\   
20111214184   & 55910.5166 & 20& 46&  11.29 & 123.7 & -205.5 &---   &  1.032	&   1.71 &  1.72  &  1.43 & -9.0  &  \\    
20111214185   & 55910.5308 & 20& 33&  11.10 & 122.6 & -199.6 &---   &  0.965	&   1.72 &  1.72  &  1.38 & -5.0  &  \\    
20111214186   & 55910.5449 & 20& 35&   9.52 & 126.1 & -201.1 &---   &  0.961	&   1.71 &  1.71  &  1.44 & -21.1 &  \\    
20111214187   & 55910.5591 & 20& 36&  11.50 & 123.9 & -207.5 &---   &  0.915	&   1.73 &  1.72  &  1.39 & -20.0 &  \\    
20120707126   & 56115.5039 & 20& 32&  10.34 & 150.8 & -185.8 &-34.0 &  0.870	&   1.83 &  1.86  &  1.28 & -24.9 &  \\    
20120707163   & 56116.5151 & 20& 34&  10.25 & 155.6 & -183.8 &-23.9 &  0.926	&   1.79 &  1.84  &  1.19 & -2.3  &  \\    
20120707164   & 56116.5306 & 20& 53&  10.84 & 153.4 & -184.9 &-23.8 &  0.928	&   1.80 &  1.86  &  1.19 & 0.6   &  \\    
20120708046   & 56117.5046 & 20& 49&  10.50 & 137.1 & -193.3 &-34.4 &  0.789	&   1.72 &  1.89  &  1.23 & -18.3 &  \\    
20120902346   & 56173.4891 & 20& 32&  10.91 & 133.2 & -185.2 &-13.4 &  1.178	&   1.74 &  1.77  &  1.24 & ---   &  \\    
20120902347   & 56173.5032 & 20& 26&  10.30 & 128.7 & -184.9 &---   &  1.156	&   1.74 &  1.79  &  1.25 & ---   &  \\    
20120928444   & 56198.5006 & 20& 63&  10.13 & 141.3 & -198.1 &-30.4 &  0.947	&   1.74 &  1.74  &  1.19 & -4.3  &  \\    
20120928445   & 56198.5156 & 20& 70&  10.12 & 137.6 & -193.2 &-30.6 &  0.936	&   1.71 &  1.75  &  1.18 & -3.9  &  \\    
2013010357    & 56296.2614 & 20& 35&   9.30 & 132.2 & -200.9 &-65.1 &  0.739	&   1.66 &  1.72  &  1.18 & -53.7 &  \\    
2013010358    & 56296.2755 & 20& 37&   9.01 & 133.3 & -206.0 &-69.4 &  0.769	&   1.64 &  1.73  &  1.18 & -32.8 &  \\    
& \\		                							     
\hline                  		 
\end{tabular}	        								      
\end{table*}	        								      
%
%


\begin{thebibliography}{}
  \bibitem[Abbasi et al.(2012)]{2012ApJ...748..118A} Abbasi, R., Abdou, Y., Abu-Zayyad, T., et al.\ 2012, \apj, 748, 118 
  \bibitem[Abdo et al.(2009)]{2009ApJ...701L.123A} Abdo, A.~A., Ackermann, M., Ajello, M., et al.\ 2009, \apjl, 701, L123 
  \bibitem[Acciari et al.(2008)]{2008ApJ...679.1427A} Acciari, V.~A., Beilicke, M., Blaylock, G., et al.\ 2008, \apj, 679, 1427 
  \bibitem[Albert et al.(2006)]{2006Sci...312.1771A} Albert, J., Aliu, E., Anderhub, H., et al.\ 2006, Science, 312, 1771 
  \bibitem[Albert et al.(2008)]{2008ApJ...684.1351A} Albert, J., Aliu, E., Anderhub, H., et al.\ 2008, \apj, 684, 1351 
  \bibitem[Apparao(2001)]{2001A&A...371..672A} Apparao, K.~M.~V.\ 2001, \aap, 371, 672 
  \bibitem[Aragona et al.(2009)]{2009ApJ...698..514A}     Aragona, C., McSwain,  M.~V., Grundstrom, E.~D., et al.\ 2009, \apj, 698, 514 
  \bibitem[Balona(2000)]{2000ASPC..214....1B} Balona, L.~A.\ 2000, IAU Colloq.~175: The Be Phenomenon in Early-Type Stars, 214, 1 
  \bibitem[Bignami et al.(1981)]{1981ApJ...247L..85B} Bignami, G.~F., Caraveo, P.~A., Lamb, R.~C., Markert, T.~H., \& Paul, J.~A.\ 1981, \apjl, 247, L85 
  \bibitem[Bildsten et al.(1997)]{1997ApJS..113..367B} Bildsten, L.,Chakrabarty, D., Chiu, J., et al.\ 1997, \apjs, 113, 367 
  \bibitem[Blay et al.(2006)]{2006A&A...446.1095B} Blay, P., Negueruela, I., Reig, P., et al.\ 2006, \aap, 446, 1095 
  \bibitem[Ca{\~n}ellas et al.(2012)]{2012A&A...543A.122C} Ca{\~n}ellas, A., Joshi, B.~C., Paredes, J.~M., et al.\ 2012, \aap, 543, A122  
  \bibitem[Casares et al.(2012)]{2012MNRAS.421.1103C} Casares, J., Rib{\'o}, M., Ribas, I., et al.\ 2012a, \mnras, 421, 1103 
  \bibitem[Casares et al.(2012)]{2012MNRAS.426..796C} Casares, J., Rib{\'o},  M., Ribas, I., et al.\ 2012b, \mnras, 426, 796 
  \bibitem[Chauville et al.(2001)]{2001A&A...378..861C} Chauville, J., Zorec, J., Ballereau, D., et al.\ 2001, \aap, 378, 861 
  \bibitem[Chernyakova et al.(2012)]{2012ApJ...747L..29C} Chernyakova, M., Neronov, A., Molkov, S., et al.\ 2012, \apjl, 747, L29 
  \bibitem[Coe et al.(2006)]{2006MNRAS.368..447C} Coe, M.~J., Reig, P., McBride, V.~A., Galache, J.~L., \& Fabregat, J.\ 2006, \mnras, 368, 447 
  \bibitem[Corbet(1986)]{1986MNRAS.220.1047C} Corbet, R.~H.~D.\ 1986, \mnras, 220, 1047 
  \bibitem[Demircan \& Kahraman(1991)]{1991Ap&SS.181..313D} Demircan, O., \& Kahraman, G.\ 1991, \apss, 181, 313 
  \bibitem[Dhawan et al.(2006)]{2006smqw.confE..52D} Dhawan, V., Mioduszewski, A., \& Rupen, M.\ 2006, VI Microquasar Workshop: Microquasars and Beyond,  
  \bibitem[Eggleton(1983)]{1983ApJ...268..368E} Eggleton, P.~P.\ 1983, \apj, 268, 368 
  \bibitem[Gregory \& Taylor(1978)]{1978Natur.272..704G} Gregory, P.~C., \& Taylor, A.~R.\ 1978, \nat, 272, 704 
  \bibitem[Gregory et al.(1979)]{1979AJ.....84.1030G} Gregory, P.~C., Taylor, A.~R., Crampton, D., et al.\ 1979, \aj, 84, 1030 
  \bibitem[Gregory et al.(1989)]{1989ApJ...339.1054G} Gregory, P.~C., Xu, H.-J., Backhouse, C.~J., \& Reid, A.\ 1989, \apj, 339, 1054 
  \bibitem[Gregory(2002)]{2002ApJ...575..427G} Gregory, P.~C.\ 2002, \apj, 575, 427 
  \bibitem[Greiner \& Rau(2001)]{2001A&A...375..145G} Greiner, J., \& Rau, A.\ 2001, \aap, 375, 145 
  \bibitem[Grundstrom et al.(2007)]{2007ApJ...656..437G} Grundstrom, E.~D., Caballero-Nieves, S.~M., Gies, D.~R., et al.\ 2007a, \apj, 656, 437 
  \bibitem[Grundstrom et al.(2007)]{2007ApJ...660.1398G} Grundstrom, E.~D., Boyajian, T.~S., Finch, C., et al.\ 2007b, \apj, 660, 1398 
  \bibitem[Hadasch et al.(2012)]{2012ApJ...749...54H} Hadasch, D., Torres, D.~F., Tanaka, T., et al.\ 2012, \apj, 749, 54 
  \bibitem[Hanuschik et al.(1988)]{1988A&A...189..147H} Hanuschik, R.~W., Kozok, J.~R., \& Kaiser, D.\ 1988, \aap, 189, 147
  \bibitem[Hanuschik(1989)]{1989Ap&SS.161...61H} Hanuschik, R.~W.\ 1989, \apss, 161, 61 
  \bibitem[Hanuschik(1996)]{1996A&A...308..170H} Hanuschik, R.~W.\ 1996, \aap, 308, 170 
  \bibitem[Hanuschik(2000)]{2000ASPC..214..518H} Hanuschik, R.~W.\ 2000, IAU Colloq.~175: The Be Phenomenon in Early-Type Stars, 214, 518 
  \bibitem[Hermsen et al.(1977)]{1977Natur.269..494H} Hermsen, W., Swanenburg, B.~N., Bignami, G.~F., et al.\ 1977, \nat, 269, 494 
  \bibitem[Hummel \& Dachs(1992)]{1992A&A...262L..17H} Hummel, W., \& Dachs, J.\ 1992, \aap, 262, L17 
  \bibitem[Haigh et al.(2000)]{2000ASPC..214..735H} Haigh, N.~J., Coe, M.~J., Steele, I.~A., \& Fabregat, J.\ 2000, IAU Colloq.~175: The Be Phenomenon in Early-Type Stars, 214, 735 
  \bibitem[Haigh et al.(2004)]{2004MNRAS.350.1457H} Haigh, N.~J., Coe, M.~J., \& Fabregat, J.\ 2004, \mnras, 350, 1457 
  \bibitem[Harrison et al.(2000)]{2000ApJ...528..454H} Harrison, F.~A., Ray, P.~S., Leahy, D.~A., Waltman, E.~B., \& Pooley, G.~G.\ 2000, \apj, 528, 454 
  \bibitem[Huang(1972)]{1972ApJ...171..549H} Huang S.-S., 1972, ApJ 171, 549
  \bibitem[Hubert \& Floquet(1998)]{1998A&A...335..565H} Hubert, A.~M., \& Floquet, M.\ 1998, \aap, 335, 565 
  \bibitem[Hummel \& Vrancken(2000)]{2000A&A...359.1075H} Hummel, W., \& Vrancken, M.\ 2000, \aap, 359, 1075 
  \bibitem[Hummel \& Dachs(1992)]{1992A&A...262L..17H} Hummel, W., \& Dachs, J.\ 1992, \aap, 262, L17  \bibitem[Hutchings \& Crampton(1981)]{1981PASP...93..486H} Hutchings, J.~B., \& Crampton, D.\ 1981, \pasp, 93, 486 
  \bibitem[Leahy(2001)]{2001ICRC....6.2524L} Leahy, D.~A.\ 2001, International Cosmic Ray Conference, 6, 2524 
  \bibitem[Li et al.(2012)]{2012ApJ...744L..13L} Li, J., Torres, D.~F., Zhang, S., et al.\ 2012, \apjl, 744, L13 
  \bibitem[Liu \& Yan(2005)]{2005NewA...11..130L} Liu, Q.~Z., \& Yan, J.~Z.\ 2005, \na, 11, 130 
  \bibitem[Massi et al.(2001)]{2001A&A...376..217M} Massi, M., Rib{\'o}, M., Paredes, J.~M., Peracaula, M., \& Estalella, R.\ 2001, \aap, 376, 217 
  \bibitem[Massi et al.(2004)]{2004A&A...414L...1M} Massi, M., Rib{\'o}, M., Paredes, J.~M., et al.\ 2004, \aap, 414, L1 
  \bibitem[Massi \& Zimmermann(2010)]{2010A&A...515A..82M} Massi, M., \& Zimmermann, L.\ 2010, \aap, 515, A82 
  \bibitem[Massi et al.(2012)]{2012A&A...540A.142M} Massi, M., Ros, E., \& Zimmermann, L.\ 2012, \aap, 540, A142 
  \bibitem[McSwain et al.(2010)]{2010ApJ...724..379M} McSwain, M.~V., Grundstrom, E.~D., Gies, D.~R., \& Ray, P.~S.\ 2010, \apj, 724, 379 
  \bibitem[McSwain et al.(2011)]{2011ApJ...738..105M} McSwain, M.~V., Ray, P.~S., Ransom, S.~M., et al.\ 2011, \apj, 738, 105 
  \bibitem[Mendelson \& Mazeh(1994)]{1994MNRAS.267....1M} Mendelson, H., \& Mazeh, T.\ 1994, \mnras, 267, 1 
  \bibitem[Mirabel et al.(2004)]{2004A&A...422L..29M} Mirabel, I.~F., Rodrigues, I., \& Liu, Q.~Z.\ 2004, \aap, 422, L29
  \bibitem[Okazaki \& Negueruela(2001)]{2001A&A...377..161O} Okazaki, A.~T., \& Negueruela, I.\ 2001, \aap, 377, 161 
  \bibitem[Paredes et al.(1997)]{1997A&A...320L..25P} Paredes, J.~M., Mart\'{\i}, J., Peracaula, M., \& Ribo, M.\ 1997, \aap, 320, L25 
  \bibitem[Paredes (1987)]{Paredes.PhD}  Paredes, J. M., 1987, PhD Thesis, Universitat de Barcelona 
  \bibitem[Paredes et al.(1994)]{1994A&A...288..519P} Paredes, J.~M., Marziani, P., Mart\'{\i}, J., et al.\ 1994, \aap, 288, 519 
  \bibitem[Paredes et al.(2013)]{2013APh....43..301P} Paredes,  J.~M.,Bednarek, W., Bordas, P., et al.\ 2013, APh, 43, 4301
  \bibitem[Peters et al.(2013)]{2013ApJ...765....2P} Peters, G.~J., Pewett, T.~D., Gies, D.~R., Touhami, Y.~N., \& Grundstrom, E.~D.\ 2013, \apj, 765, 2 
  \bibitem[Porter(1996)]{1996MNRAS.280L..31P} Porter, J.~M.\ 1996, \mnras, 280, L31 
  \bibitem[Porter \& Rivinius(2003)]{2003PASP..115.1153P} Porter, J.~M., \& Rivinius, T.\ 2003, \pasp, 115, 1153 
  \bibitem[Punsly(1999)]{1999ApJ...519..336P} Punsly, B.\ 1999, \apj, 519, 336 
  \bibitem[Rea et al.(2010)]{2010MNRAS.405.2206R} Rea, N., Torres, D.~F., van der Klis, M., et al.\ 2010, \mnras, 405, 2206 
  \bibitem[Reid \& Parker(2012)]{2012MNRAS.425..355R} Reid, W.~A., \& Parker, Q.~A.\ 2012, \mnras, 425, 355 
  \bibitem[Reig et al.(1997)]{1997A&A...322..193R} Reig, P., Fabregat, J., \& Coe, M.~J.\ 1997, \aap, 322, 193 
  \bibitem[Reig et al.(2005)]{2005A&A...440.1079R} Reig, P., Negueruela, I., Fabregat, J., Chato, R., \& Coe, M.~J.\ 2005, \aap, 440, 1079 
  \bibitem[Reig(2011)]{2011Ap&SS.332....1R} Reig, P.\ 2011, \apss, 332, 1 
  \bibitem[Rivinius et al.(2006)]{2006A&A...459..137R} Rivinius, T., {\v S}tefl, S., \& Baade, D.\ 2006, \aap, 459, 137 
  \bibitem[Roberts et al.(1987)]{1987AJ.....93..968R} Roberts, D.~H., Lehar, J., \& Dreher, J.~W.\ 1987, \aj, 93, 968 
  \bibitem[Roche et al.(1997)]{1997A&A...322..139R} Roche, P., Larionov, V., Tarasov, A.~E., et al.\ 1997, \aap, 322, 139 
  \bibitem[Romero et al.(2007)]{2007A&A...474...15R} Romero, G.~E., Okazaki, A.~T., Orellana, M., \& Owocki, S.~P.\ 2007, \aap, 474, 15 
  \bibitem[Slettebak(1988)]{1988PASP..100..770S} Slettebak, A.\ 1988, \pasp, 100, 770 
  \bibitem[Smith et al.(2009)]{2009ApJ...693.1621S} Smith, A., Kaaret, P., Holder, J., et al.\ 2009, \apj, 693, 1621 
  \bibitem[Steele et al.(1996)]{1996A&AS..120C.213S} Steele, I.~A., Coe, M.~J., Fabregat, J., et al.\ 1996, \aaps, 120, 213 
  \bibitem[Stellingwerf(1978)]{1978ApJ...224..953S} Stellingwerf, R.~F.\ 1978, \apj, 224, 953 
  \bibitem[Stoyanov \& Zamanov(2009)]{2009AN....330..727S} Stoyanov, K.~A., \& Zamanov, R.~K.\ 2009, Astronomische Nachrichten, 330, 727 
  \bibitem[Sung \& Lee(1995)]{1995JKAS...28..119S} Sung, H., \& Lee, S.-W.\ 1995, Journal of Korean Astronomical Society, 28, 119 
  \bibitem[Tarasov et al.(2003)]{2003A&A...402..237T} Tarasov, A.~E., Brocksopp, C., \& Lyuty, V.~M.\ 2003, \aap, 402, 237 
  \bibitem[Taylor et al.(1992)]{1992ApJ...395..268T} Taylor, A.~R., Kenny, H.~T., Spencer, R.~E., \& Tzioumis, A.\ 1992, \apj, 395, 268 
  \bibitem[Torres et al.(2010)]{2010ApJ...719L.104T} Torres, D.~F., Zhang, S., Li, J., et al.\ 2010, \apjl, 719, L104 
  \bibitem[Underhill \& Doazan(1982)]{1982bsww.book.....U} Underhill, A., \& Doazan, V.\ 1982,  B Stars with and without emission lines (NASA SP-456)
  \bibitem[Zaitseva \& Borisov(2003)]{2003AstL...29..188Z} Zaitseva, G.~V., \& Borisov, G.~V.\ 2003, Astronomy Letters, 29, 188 
  \bibitem[Zamanov(1995)]{1995MNRAS.272..308Z} Zamanov, R.~K.\ 1995, \mnras, 272, 308 
  \bibitem[Zamanov \& Mart{\'{\i}}(2000)]{2000A&A...358L..55Z} Zamanov, R., \& Mart{\'{\i}}, J.\ 2000a, \aap, 358, L55 
  \bibitem[Zamanov et al.(1999)]{1999A&A...351..543Z} Zamanov, R.~K., Mart{\'{\i}}, J., Paredes, J.~M., et al.\ 1999, \aap, 351, 543 
  \bibitem[Zamanov et al.(2001)]{2001cnoc.conf...50Z} Zamanov, R., Marti, J., 
   \& Marziani, P.\ 2001, The Second National Conference on Astrophysics of Compact Objects, 50 (astro-ph/0110114)


\end{thebibliography}
\end{document}